%% file: main.tex
\documentclass{vgtc}                          

\ifpdf
  \pdfoutput=1\relax                   
  \usepackage{graphicx}                
  \DeclareGraphicsExtensions{.pdf,.png,.jpg,.jpeg} 
\else
  \usepackage{graphicx}                
  \DeclareGraphicsExtensions{.eps}     
\fi%

\graphicspath{{figs/}}

\usepackage{microtype}                 
\PassOptionsToPackage{warn}{textcomp}  
\usepackage{textcomp}                  
\usepackage{mathptmx}                  
\usepackage{times}                     
\usepackage{cite}                      
\usepackage{tabu}                      
\usepackage{booktabs}                  
\usepackage{graphics} 
\usepackage{epsfig} 
\usepackage{amsmath} 
\usepackage{amssymb}  
\usepackage{color}
\usepackage{subfig}
\usepackage{cite}
\usepackage{siunitx}
\usepackage{soul}
\usepackage{tikz}
\usepackage{xspace}
\usepackage{url}

\usepackage{eso-pic}

\AddToShipoutPictureBG*{%
\begin{tikzpicture}[remember picture, overlay]
      \node[minimum width=4in,font=\itshape] at ([yshift=-1cm]current page.north)  {This is the extended version of the article published at IEEE Conference on Virtual Reality and 3D User Interfaces (IEEE VR), 2022};
    \end{tikzpicture}%

}

\onlineid{0}

\vgtccategory{Research}

\vgtcinsertpkg

\title{Real-Time Gaze Tracking with Event-Driven Eye Segmentation}

\author{Yu Feng\thanks{e-mail: yfeng28@ur.rochester.edu}\\ %
        \scriptsize University of Rochester %
\and Nathan Goulding-Hotta\thanks{e-mail: nrgh@fb.com}\\ %
    \scriptsize Reality Labs Research %
\and Asif Khan\thanks{e-mail: aikhan@fb.com}\\ %
    \scriptsize Reality Labs Research %
\and Hans Reyserhove\thanks{Now at Impinj, e-mail: hans.reyserhove@impinj.com}\\ %
    \scriptsize Reality Labs Research %
\and Yuhao Zhu\thanks{e-mail: yzhu@rochester.edu}\\ %
    \parbox{1.4in}{\scriptsize \centering University of Rochester}}

\teaser{
  \centering
  \includegraphics[width=\linewidth]{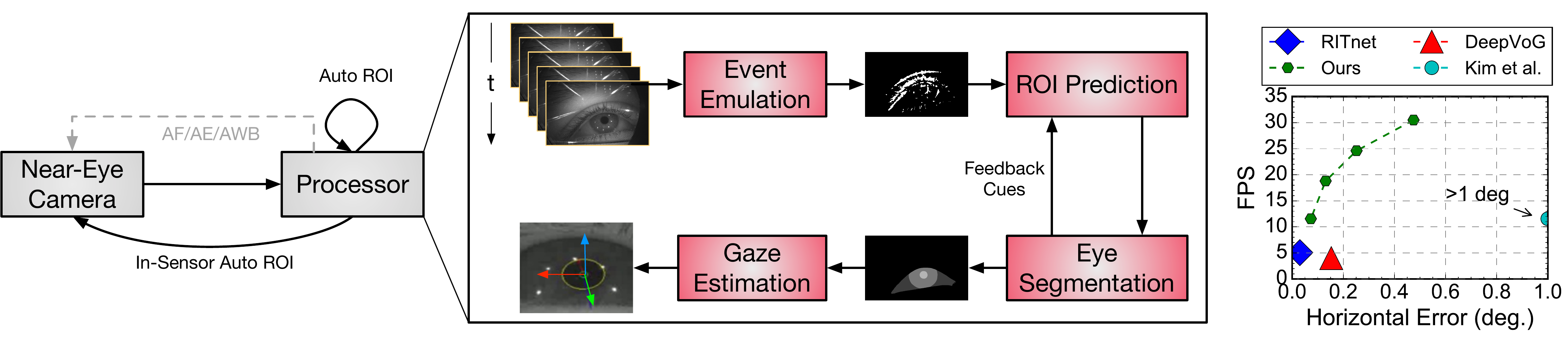}
  \caption{\textbf{Left}: Workflow of our gaze tracking algorithm. The algorithm provides an Auto~ROI mode (akin to the conventional ``3A'' algorithms in a camera) by emulating an event camera in software and using the events to predict the Region of Interest (ROI). Performing Auto~ROI in emerging compute-capable image sensors further reduces the data transmission energy. \textbf{Right}: Our algorithm achieves sub-\ang{0.5} gaze accuracies (both horizontal and vertical directions) using only 30K parameters, an order of magnitude smaller than other algorithms. Marker sizes in the figure are proportional to the number of parameters. The algorithm operates at over 30 Hz on a mobile Jetson Xavier processor.}
  \label{fig:teaser}
}

\input{macro}

\input{abstract}

\CCScatlist{
\CCScatTwelve{Computing methodologies}{Computer graphics}{Graphics systems and interfaces}{Mixed/augmented reality};
 \CCScatTwelve{Computing methodologies}{Computer graphics}{Shape modeling}{Shape analysis}
}

\begin{document}



\maketitle

\input{intro}
\input{bg}
\input{framework}
\input{setup}
\input{eval}
\input{conc}

\bibliographystyle{abbrv-doi}

\bibliography{refs}
\end{document}

%% file: macro.tex
\ifx\figurename\undefined \def\figurename{Figure}\fi
\renewcommand{\figurename}{Fig.}
\renewcommand{\paragraph}[1]{\textbf{#1}~~}

\newcommand{\Sect}[1]{Section~\ref{#1}}
\newcommand{\Fig}[1]{Figure~\ref{#1}}
\newcommand{\Tbl}[1]{Table~\ref{#1}}
\newcommand{\Eqn}[1]{Equation~\ref{#1}}

\newcommand{\specialcell}[2][c]{\begin{tabular}[#1]{@{}l@{}}#2\end{tabular}}

\newcommand{\no}[1]{#1}

\newcommand{\RNum}[1]{\uppercase\expandafter{\romannumeral #1\relax}}


%% file: abstract.tex
\abstract{Gaze tracking is increasingly becoming an essential component in Augmented and Virtual Reality. Modern gaze tracking algorithms are heavyweight; they operate at most 5 Hz on mobile processors despite that near-eye cameras comfortably operate at a real-time rate ($>$ 30 Hz). This paper presents a real-time eye tracking algorithm that, on average, operates at 30 Hz on a mobile processor, achieves \ang{0.1}--\ang{0.5} gaze accuracies, all the while requiring only 30K parameters, one to two orders of magnitude smaller than state-of-the-art eye tracking algorithms. The crux of our algorithm is an Auto~ROI mode, which continuously predicts the Regions of Interest (ROIs) of near-eye images and judiciously processes only the ROIs for gaze estimation. To that end, we introduce a novel, lightweight ROI prediction algorithm by emulating an event camera. We discuss how a software emulation of events enables accurate ROI prediction without requiring special hardware. We also show that our algorithm is amenable to emerging compute-capable image sensors. By pushing the Auto~ROI mode into the sensor, akin to today's ``3A'' algorithms, we further reduce the data transmission energy by over 50\%. The code of our paper is available at \color{blue}{\url{https://github.com/horizon-research/edgaze}}.
}

%% file: intro.tex
\section{Introduction}
\label{sec:intro}

Gaze tracking is critical to many fields~\cite{mania2021gaze}, such as medicine~\cite{brunye2019review, borys2017eye}, human-machine interaction~\cite{strandvall2009eye, jacob2003eye, chandra2015eye, mathur2021dynamic}, psychology~\cite{mele2012gaze}, augmented reality (AR)~\cite{itoh2014interaction, plopski2016automated}, and virtual reality (VR)~\cite{hu2020gaze, clay2019eye, whitmire2016eyecontact, patney2016perceptually, zhang2017eye}. Due to its importance, there has been much research improving the tracking accuracy, both using conventional infrared/RGB cameras~\cite{chaudhary2019ritnet,kothari2021ellseg, kim2019eye, yiu2019deepvog, jianfeng2014eye} and emerging event cameras~\cite{angelopoulos2020event}.

The speed of gaze tracking is critical. Gaze tracking, in many cases, directly interfaces with users, and is often just the first stage of an end-to-end application. Studies have shown that a tracking speed of at least 30 Hz is required for real-time user-facing applications such as VR~\cite{al2013enhanced}. Today's eye tracking systems, however, are generally an order of magnitude slower than the requirement. For instance, on Nvidia's Jetson Xavier, a powerful mobile computing platform, two state-of-the-art algorithms, RITNet~\cite{chaudhary2019ritnet} and DeepVOG~\cite{yiu2019deepvog}, operate at a mere 5 Hz (while achieving a sub-\ang{0.5} gaze error).

This paper demonstrates a gaze tracking system that achieves 30 Hz tracking speed on a mobile GPU with a sub-\ang{0.5} gaze accuracy. Most prior eye tracking algorithms work on a frame-by-frame basis. However, eye tracking in virtually all use-cases processes continuous frames, which exposes temporal information that is often ignored. Our system leverages this temporal correlation for real-time tracking.

\paragraph{Event-Driven Auto ROI} In particular, we use the temporal correlation across frames to predict the Region of Interest (ROI) in the current frame. We perform processing only in the ROIs whenever possible, falling back to full-resolution processing very rarely (0.6\% of the time). The ROIs usually contain only 18\%-32\% of the pixels compared to full resolution, greatly improving the execution speed.

Our contribution here is a novel ROI prediction algorithm specialized for eye tracking. In particular, we emulate an event camera to extract the temporal correlation across frames, which then guides a lightweight deep neural network (DNN) to predict the ROI. The ROI prediction network essentially gives rise to an \textit{Auto ROI mode} for eye tracking, akin to the conventional 3A algorithms for photography (auto focus, auto white balance, and auto exposure). Similar to the 3A algorithms, this ``fourth A'' mode operates in a feedback-driven fashion, where we use information extracted from the current frame (e.g., events, edges, ROI) to predict the ROI of future frames.

\paragraph{Why Events?} Event cameras report only the changes in pixel brightness level, a.k.a., events. We use events for ROI prediction for two reasons. First, events inherently encode the temporal changes, and thus provide natural cues to ROI prediction. Second, events are naturally sparse and can be efficiently encoded, enabling a small neural network with a low overhead in ROI prediction. Computation for the ROI prediction must be lightweight, in order to not offset the gains from processing only ROI images.

\paragraph{Why Not an Event Camera?}  We do not directly use an event camera for two main reasons. First, event cameras are not readily available in many real-world systems (e.g., VR) and usually require dedicated downstream algorithms, which limit their applicability. Second, although event cameras can operate at a frequency as high as 10K Hz, useful for capturing precise eye movement, such a high speed is usually an overkill for many consumer eye-tracking applications, e.g., user interface navigation in VR, where the target speed is ``only'' 30 Hz, easily achievable with conventional cameras.

\paragraph{Efficiency-Oriented Segmentation} Coupled with the Auto~ROI mode, we propose a lightweight neural network-based algorithm to predict the gaze from an ROI. Like much of the recent work~\cite{chaudhary2019ritnet, yiu2019deepvog, kothari2021ellseg, kim2019eye}, we use a model-based approach, where we first parameterize an eye image through eye segmentation, which is then used to estimate the gaze through fitting a geometric eye model.

We propose a new segmentation network designed with efficiency in mind. We show that depth-wise separable convolution with wider channels provides a segmentation backbone that balances accuracy and compute complexity. Our network requires 13--93$\times$ fewer parameters (30K in total) and 20--31$\times$ fewer arithmetic operations than existing DNN algorithms (RITnet~\cite{chaudhary2019ritnet} and DeepVOG~\cite{yiu2019deepvog}).

\no{\paragraph{Reducing Data Transmission} Apart from delivering real-time tracking on off-the-shelf mobile platforms today, our system has the potential to reduce data transmission overhead, both between the image sensor and the back-end processor and between the processor and memory. Data transmission overhead is a major target for optimization, as the per-byte energy consumption of data transmission is almost three orders of magnitude higher than that of computation~\cite{liu2019intelligent}.

In particular, we show that our ROI prediction network, due to its small memory footprint and lightweight compute complexity, is amenable to execution on compute-capable image sensors, allowing only the ROI pixels or a small amount of metadata, as opposed to full-resolution images, to be transmitted out of the sensor. Using an industry-grade analytical model~\cite{liu2019intelligent}, we show that in-sensor Auto~ROI provides 2--4$\times$ reduction in the data transmission energy.}

Our dataset and code will be released. Our contributions are:
\begin{itemize}
    \item We introduce \textit{event-driven Auto~ROI} for eye tracking, a novel event-driven ROI prediction algorithm for eye tracking using software-emulated events and temporal feedback.
    \item We propose an accurate eye segmentation neural network co-trained with ROI prediction;  the segmentation network is an order of magnitude simpler compared to prior algorithms.
    \no{\item We show how our system can be integrated into emerging compute-capable image sensors to reduce the sensor data transmission energy for eye tracking.}
\end{itemize}

%% file: bg.tex
\section{Related Work} 
\label{sec:bg}

\begin{figure}[t]
\centering
\includegraphics[width=\columnwidth]{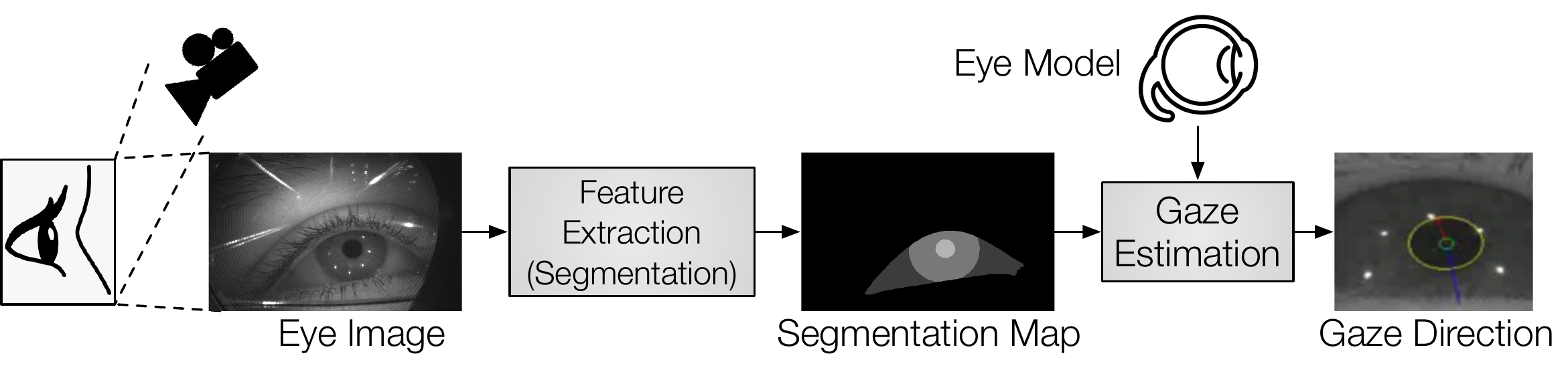}
\caption{A typical model-based eye tracking pipeline, which consists of a feature extraction stage and a gaze estimation stage. Feature extraction is usually done through eye segmentation. Gaze estimation is much more lightweight than feature extraction.}
\label{fig:overview}
\end{figure}

\paragraph{Eye Tracking Algorithms} Eye tracking methods fall into two main categories: model-based methods and appearance-based methods~\cite{hansen2009eye, zhang2019evaluation}. Appearance-based methods directly learn a mapping from an eye image to the gaze directions~\cite{zhang2017mpiigaze, zhang2015appearance, wood2016learning, lu2014adaptive}. We focus on the model-based method, which predicts gazes using a physiology-inspired eye model. Model-based methods are generally regarded as providing better accuracy than appearance-based methods~\cite{zhang2019evaluation}.

\Fig{fig:overview} illustrates the pipeline of a typical model-based method~\cite{yiu2019deepvog, li2018etracker}, which consists of two stages: 1) eye feature extraction and 2) gaze estimation. Feature extraction algorithms take near-eye images  as inputs and extract salient features of the eye. The features represent a parameterization of the eye. The parameters are then used to fit a geometric eye model to estimate the gaze. The eye model is a physics-based 3D model typically pre-constructed and calibrated from experimental data~\cite{yiu2019deepvog, hennessey2006single, chen20083d, zhu2005eye, li2018etracker}. 

Compared to gaze estimation, feature extraction is much more time consuming. Our experiments show that feature extraction contributes to 85\% of the total execution time across a range of different algorithms, and is thus a prime target for optimization. We introduce a lightweight neural network to extract features through segmentation, which reduces the computation complexity by an order of magnitude compared to state-of-the-art networks.

\paragraph{Feature Extraction} While historically hand-crafted, geometric features are popular~\cite{hansen2005eye, fuhl2016else, swirski2012robust, itohtracker}, DNN algorithms have recently been shown to be more robust and accurate~\cite{chaudhary2019ritnet,yiu2019deepvog, kothari2021ellseg, kim2019eye, fuhl2016pupilnet, li2018etracker}. In particular, extracting features through eye segmentation~\cite{chaudhary2019ritnet, yiu2019deepvog, kothari2021ellseg, kim2019eye} is by far the most widely used method.

An eye segmentation algorithm usually classifies the pixels in an eye image into different classes. The resulting segmentation map has the same dimensions as the eye image, and each pixel value in the map represents the class ID of that pixel. Our algorithm uses a similar approach as DeepVOG~\cite{yiu2019deepvog}, RITnet~\cite{chaudhary2019ritnet}, OpenEDS~\cite{openeds2019}, and Kim et al.~\cite{kim2019eye}, known as ``part segmentation,'' which segments the eye image into four parts: pupil, iris, sclera, and background. In contrast, Ellseg~\cite{kothari2021ellseg} and Wang et al.~\cite{wang2021edge} use elliptical segmentation, which predicts the ellipses of the pupil and iris.

Prior studies focus primarily on accuracy. In contrast, we focus on \textit{compute efficiency}, and show that tracking can be five times faster with little sacrifice in accuracy. We note that while we demonstrate our ROI prediction on part segmentation-based eye tracking, the idea applies generally to tracking algorithms using other features.

\paragraph{Event Camera} Event cameras operate each pixel independently and asynchronously~\cite{gallego2019event}. Each pixel gets activated when its intensity change surpasses a predefined threshold. The response is called an ``event'', which includes the pixel coordinates, a timestamp, and a polarity value. Because of the increased hardware complexity, the resolution of event cameras is typically lower than that of conventional cameras~\cite{gallego2019event}, but event cameras have a much higher frame rate (upward of tens of thousands of Hz) since they produce only (sparse) events occasionally rather than (dense) pixels regularly.


Event cameras appear in a range of vision and robotics tasks such as object tracking~\cite{ramesh2018long}, localization~\cite{weikersdorfer2013simultaneous}, and reconstruction~\cite{kim2016real}. Recent work has also started using event cameras for eye tracking~\cite{angelopoulos2020event, damianeye}, and is able to achieve a 10K Hz frequency. While extremely high tracking frequency is needed when capturing precise eye movement (e.g., foveated rendering), a lower frame rate provided by conventional cameras is sufficient for many eye tracking use cases, e.g., eye communication system for disability~\cite{caligari2013eye}. Instead of using event camera hardware, we emulate event camera output from our conventional sensor using software.


\paragraph{ROI Computing} ROI is widely used to reduce the overall computation and data transmission~\cite{zhu2018euphrates, feng2019asv, kong2016hypernet, ren2015faster, girshick2015fast, girshick2014rich, he2017mask, kodukula2021rhythmic, mudassar2019camera}. Classic work such as fast R-CNN~\cite{girshick2015fast} use dedicated region proposal networks that are computationally heavy. Other approaches use simple extrapolation~\cite{zhu2018euphrates, feng2019asv, mudassar2019camera}, which we find insufficient for eye tracking, because the objects (eyes) move rapidly. Many image sensors provide an ROI output mode~\cite{onseminoip1sn025ka, ov9782}, but rely on users to provide the ROI coordinates. Sony built a sensor that automatically detects an ROI to drive spatial resolution and exposure modulation~\cite{kumagai20181}.

Our contribution is a lean and accurate ROI prediction algorithm tailored to eye tracking. We show that software-emulated events, combined with edge information from previous segmentation results, can effectively predict the eye movement and, by extension, the ROI.

\no{\paragraph{In-Sensor Computing} With the advent of 3D stacking technology~\cite{loh2007processor} and the maturity of digital pixel sensors~\cite{el1999pixel, liu20204}, we are witnessing a confluence of image sensing and digital processing. Companies such as Samsung~\cite{kwon2020low}, Sony~\cite{eki20219, haruta20174, tsugawa2017pixel}, Meta~\cite{liu20204} and OmniVision~\cite{venezia20181} have demonstrated image sensors that vertically stack a pixel array, memory, and digital processor together in one sensor. The processors in today's stacked sensors are mostly dedicated image signal processors. Trends indicate, however, that these processors will evolve into full-fledged programmable cores (micro-controllers) or DNN accelerators~\cite{mukhopodhyay2018camel, amir20183d, eki20219}.

Compute-capable image sensors present a new design space to explore, wherein the central question is: What type of computation is profitable to move into the sensor? This paper presents an interesting case-study where the lightweight ROI prediction executes inside the sensor to reduce data transmission outside the sensor.}

%% file: framework.tex
\section{Event-Driven Auto ROI for Eye Tracking}
\label{sec:framework}

\begin{figure}[t]
  \centering
  \includegraphics[width=\columnwidth]{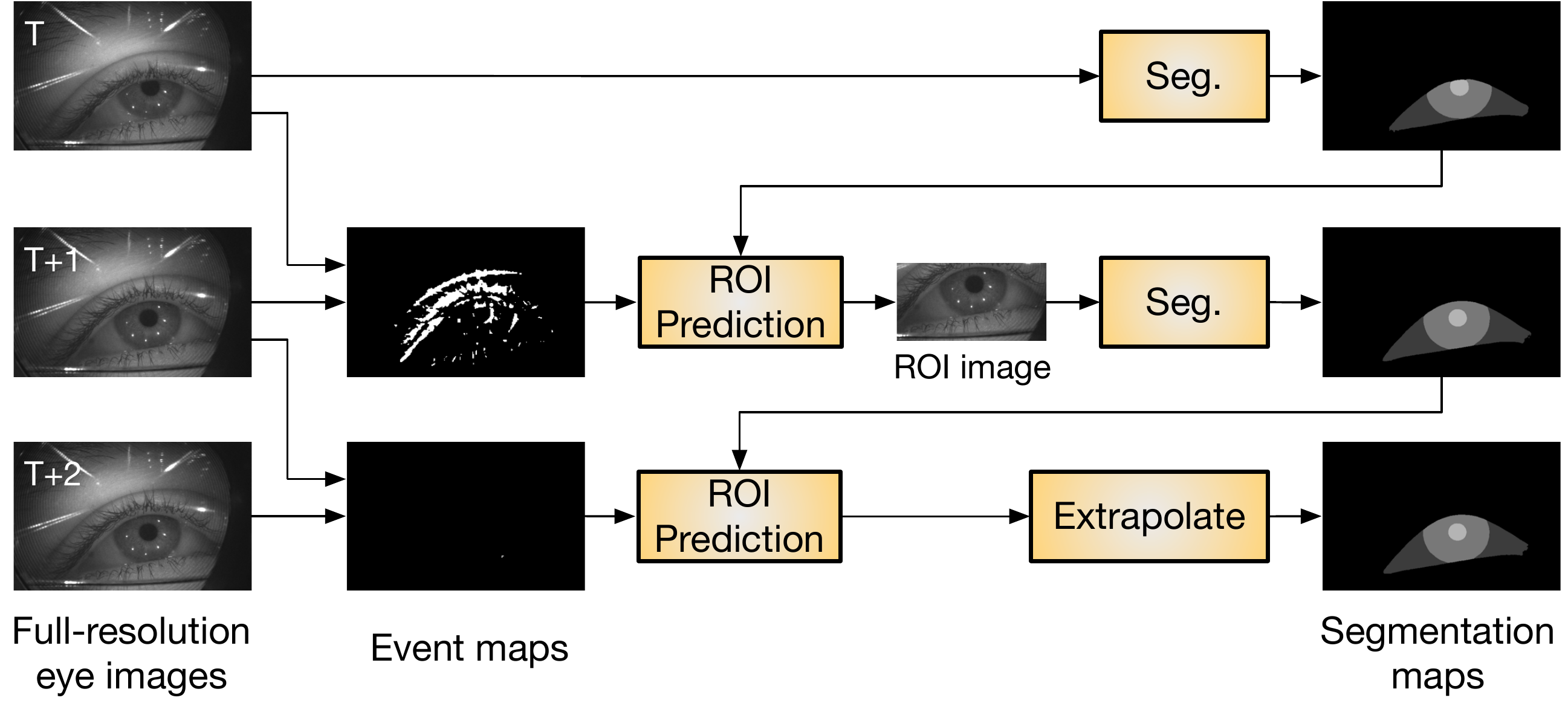}
  \caption{Overview of our event-driven eye segmentation. Time progresses from top to bottom. The segmentation results are used by a common gaze estimation algorithm~\cite{Swirski2013}, which is omitted in the figure. We generate an event map from every two consecutive eye frames; the current event map and previous segmentation map are combined to predict the ROI. If the number of events is small (e.g., shown at time ($T+2$)), we can simply extrapolate the segmentation result rather than performing a full-blown segmentation.}
  \vspace{-10pt}
  \label{fig:eye_tracking}
\end{figure}

\input{roi_fig}

Gaze estimation models rely on the geometry of foreground eye parts such as the pupil, iris, and sclera. Full-resolution eye images captured by near-eye cameras usually contain a large chunk of eye muscle/skin that is irrelevant to gaze tracking. We introduce an Auto ROI mode for eye tracking, where we predict and process only the ROI that contains the foreground eye classes needed for gaze estimation. This Auto ROI mode is coupled with a new and efficient eye segmentation network.

We first provide an overview of the algorithm (\Sect{sec:framework:overview}), followed by our feedback-driven, event-based ROI prediction algorithm (\Sect{sec:framework:roi}) and the eye segmentation network (\Sect{sec:framework:seg}).

\subsection{Overview} 
\label{sec:framework:overview}

\Fig{fig:eye_tracking} shows a high-level flow diagram of our gaze tracking algorithm. We use a model-based, two-stage algorithm for gaze tracking. This paper's contributions lie in the first stage, i.e., eye segmentation, while relying on a commonly-used gaze estimation model for the second stage~\cite{Swirski2013}. Gaze estimation is relatively more mature and is much less compute intensive that the first stage (regression vs. DNN). \Fig{fig:eye_tracking} thus omits the common gaze estimation part.

Initially at time $T$, the full-resolution eye frame is processed directly by the segmentation network to generate the segmentation map. The next frame at $(T+1)$, instead of being processed directly, generates an event map, which is of the same dimension as the original full-resolution image. Each event map pixel is a 0/1 bitmask, indicating whether the corresponding pixel in the original image has a large change in intensity. This process emulates an actual event camera with simplifications tailored to eye segmentation.

The event map provides a useful guidance to predict the ROI in the current frame. We define the ROI as the minimal bounding box that encloses the three foreground eye parts, i.e., the pupil, iris, and sclera. Our approach can also apply to other tracking algorithms that use fewer or more segments (e.g.,~\cite{yiu2019deepvog, kothari2021ellseg}).

Events \textit{alone}, however, are not robust enough. When the background eye muscles move significantly and/or when foreground eye parts move little between frames, activated events do not accurately capture the segment boundaries. To improve the robustness, we propose a feedback mechanism, where two important cues from time $T$ (previous frame) are used to augment the ROI prediction.

For cases where the entire eye moves little between frames, e.g., at time $(T+2)$ in \Fig{fig:eye_tracking}, our ROI prediction algorithm would detect the inactivities and opt to extrapolate from the previous segmentation map. Having this mode ensures ``activity-proportional'' tracking, where little tracking work is done when little activity is observed.

In most cases, only the first frame has to be processed in its full resolution. In rare ($\sim$0.02\%) cases where the predicted ROI is physically infeasible (e.g., the top-right corner is to the left of the bottom-left corner), we fall back to the full-resolution mode.


\subsection{Auto ROI with Event-Driven ROI Prediction}
\label{sec:framework:roi}


The ROI prediction algorithm has two roles: 1) predicting the ROI of the current eye frame, and 2) deciding whether the current eye frame requires going through a full-fledged segmentation algorithm or can be extrapolated from the previous segmentation map. \Fig{fig:bbox_prediction} shows the structure of our ROI prediction network.

\subsubsection{The Prediction Algorithm}
\label{sec:framework:roi:pred}

\paragraph{Intuition} The goal of ROI prediction is to filter out background pixels (eye muscles, skin, and eyelids) while leaving only the foreground eye parts, i.e., the pupil, iris, and sclera. We use events to guide the ROI prediction. The intuition is that the background parts do not move as significantly as the foreground eye parts. Thus, activated pixels in the event map mostly correspond to the foreground eye parts and provide a useful guidance to the ROI prediction.

We generate the event map by emulating the high-level process of an event camera~\cite{gallego2019event}. We first calculate the pixel-wise absolute difference, $\Delta F_{t+1}$, between two consecutive frames, $F_{t}$ and $F_{t+1}$. Next, each difference value, $\Delta F_{t}(x, y)$, is normalized by the corresponding pixel value in $F_{t}$. The normalized pixel values go through an activation function, $\Phi$, which generates an event when the normalized difference is greater than a predefined threshold value. Mathematically, generating an event map can be expressed as:

\begin{equation}
E_{t+1}(x, y) = \Phi(|F_t(x, y) - F_{t+1}(x, y)|/F_t(x, y), \sigma)    
\label{eqn:event_func}
\end{equation}

\noindent where $E_{t+1}(x, y)$ is the value at coordinate $(x, y)$ of the event map at time $(T+1)$, and $\Phi$ is the activation function which outputs 1 if the difference is greater than the threshold $\sigma$ (and 0 otherwise). The threshold is a parameter that can be tuned for a specific application or scenario. \Sect{sec:eval:sen} will show that the tracking accuracy is robust against the choice of $\sigma$, but there does exist a sweet spot. Normalizing pixel differences by the previous values mimics the log-scale absolute difference operation done by an actual event camera (i.e., $log(a)-log(b)=log(a/b)$), where the pixel values are naturally in the log scale~\cite{gallego2019event}.

\begin{figure}[t]
\centering
\includegraphics[width=\columnwidth]{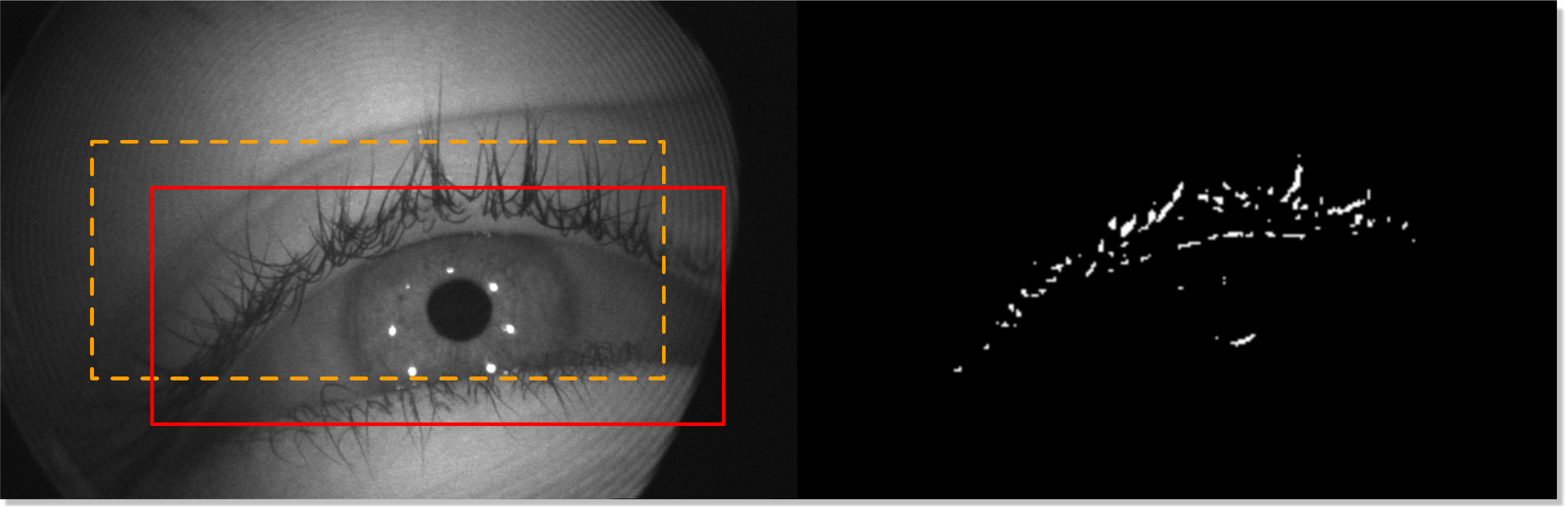}
\caption{Example of ROI misprediction using only the event map. In the left image, the solid ROI is ground truth, and the dashed ROI is the predicted ROI from using only the event map.}
\vspace{-10pt}
\label{fig:roi_cue}
\end{figure}

\input{seg_fig}

\paragraph{Two Feedback Cues} While events are largely effective, there are cases where using events alone fails. In particular, when eye muscles/eyelids move significantly and/or foreground eye parts move little, events do not accurately capture the eye boundary anymore, challenging the assumption of using events to predict ROIs.

\Fig{fig:roi_cue} shows one such example, where the left panel shows the eye frame and the right panel shows the event map (generated from the current and previous frame). The dashed-line box is the predicted ROI using only the event map, whereas the solid-line box is the ground truth ROI. In this example the upper eyelid (irrelevant to gaze estimation) moves upward with the iris (used in gaze estimation); as a result, activated events capture both the eyelids and the iris, leading to a predicted ROI significantly above the actual eye.

To cope with the issue where events in the current event map do not accurately represent the foreground eye parts, we feed the previous ROI back to the prediction algorithm. The intuition is that 
the ROI of the previous frame provides useful information to predict the ROI in the current frame due to the temporal correlation. Thus, we feed the previous ROI into the ROI prediction algorithm (a DNN), which learns to correlate the previous ROI with the current ROI.


Using the previous predicted ROI, however, has an inherent downside: the ROI prediction errors will accumulate and the predicted ROI will drift over time. To mitigate error accumulation, we use an additional cue that does \textit{not} drift over time --- the previous segmentation map. The segmentation map is predicted directly from an actual eye image from the camera; thus, the segmentation map does not drift over time. To reduce data size, we extract an edge map from the segmentation map. The edge is defined as the boundary between two different classes in the segmentation map. We use Canny edge detection~\cite{canny1986computational} to obtain an edge map from a segmentation map. Each pixel value in the edge map is a bitmask that indicates whether the corresponding pixel in the eye frame is a boundary.

\paragraph{Prediction Network} With the guidance of the event map and the two feedback cues, ROI prediction can be very lightweight. \Fig{fig:bbox_prediction} shows the network architecture, which contains three convolution (\texttt{Conv}) layers and two full-connected (\texttt{FC}) layers. The event map and edge map are concatenated first and used as the input to the \texttt{Conv} layers due to the 2D nature of the two maps. In contrast, the ROI is a $1\times 4$ vector; thus, it is concatenated with the flattened output of the \texttt{Conv} layers before entering the \texttt{FC} layers.

To reduce the overhead of the ROI prediction, we downsample the dimensions by 2 for both the event map and the edge map. We normalize the ROI to the image width/height so that each of the four coordinates in the ROI is within the [0, 1] range. We find that it is easier for the network to learn normalized values, as opposed to the absolute coordinates, which vary by camera and scenarios.

\subsubsection{Activity-Proportional Segmentation}
\label{sec:framework:roi:extra}

In cases where the entire eyes do not move across frames, detecting the inactivities and skipping segmentation all together improves the execution speed. This allows our eye tracker to be activity-proportional: no work when no activity is detected. Two issues remain: how to \textit{detect} inactivities and how to \textit{compute} an accurate segmentation map extremely fast for inactive eye frames?

Building on top of our ROI prediction algorithm, we detect inactivity by calculating the event density of the event map inside the predicted ROI. If the event density is lower than a threshold $\gamma$, the current frame is deemed inactive, in which case the ROI prediction algorithm sets the extrapolation bit, indicating that no segmentation is to be executed for the current frame. While other inactivity detection schemes are possible, we favor the simple thresholding scheme because the threshold $\gamma$ provides a useful knob to control the speed-vs-accuracy trade-off, allowing our system to potentially adapt to different application requirements and hardware capability.

We experiment with a range of extrapolation schemes, ranging from simply scaling the segments from the previous frame to using a neural network to predict how to morph the segmentation from the previous frame. We eventually settled for the simplest scheme, where the previous segmentation result is reused. In retrospect, this scheme is the most robust because it relies the least on the inactivity detection scheme, which is a simple thresholding.

\input{algorithm}

\no{
\subsection{Reducing Sensor Data Transmission}
\label{sec:framework:sensor}

So far we have focused on improving the compute efficiency of the eye tracking algorithm. However, the data transmission overhead, both between the image sensor and the processor and between the processor and memory, is non-trivial. A recent study from Facebook shows that the average energy of data transmission per byte is about 800 times higher than the computation energy per byte~\cite{liu2019intelligent}. Similar observations are made in other recent studies~\cite{kodukula2021rhythmic, han2016eie}.

The Auto ROI capability of our eye tracking algorithm provides an opportunity to reduce the data transmission overhead. Our idea is to execute part of the algorithm inside the image sensor to reduce the data transmission \textit{volume} and thereby the transmission energy. In-sensor computing is possible as image sensors integrate processing capabilities with the conventional pixel array in one sensor, sometimes in one single chip~\cite{chai2020sensor, kwon2020low, haruta20174}. A recent example is Sony's IMX 500 image sensor that provides dedicated memories and a digital signal processor for executing DNNs~\cite{imx500, eki20219}.

\begin{table}[ht]
\caption{Total number of Floating Point Operations (FLOPs) and the output data size of the four major components in our algorithm for a $640\times400$ pixel grayscale input. For this estimate we assume the ROI image is one-third the size of the full resolution, and that half of the frames are extrapolated. We show the full-resolution image size for reference. Recall that data transfer consumes 800 times more energy than computation per byte~\cite{liu2019intelligent}.}
\centering
\resizebox{1\columnwidth}{!}{
\renewcommand*{\arraystretch}{1}
\renewcommand*{\tabcolsep}{2pt}
\begin{tabular}{c c c c c | c}
\toprule[0.15em]
\textbf{Stages} & \textbf{\specialcell{~Event Map\\ Generation}} & \textbf{\specialcell{~Edge Map\\ Generation}} & \textbf{\specialcell{Prediction\\~Network}}  & \textbf{\specialcell{~~~~Seg.\\Network}} & \textbf{\specialcell{Full-res\\~Image}}\\
\midrule[0.05em]
FLOPs (Million) & 0.3 & 1.9 & 55.4 & 2641.6 & NA \\
Output Data (KB) & 7.8 & 7.8 & 41.7 & 62.5 & 250 \\
\bottomrule[0.15em]
\end{tabular}
}
\label{table:pipeline_mac}
\end{table}

While conceptually simple, reaping the reward of in-sensor computing is not trivial. Computation inside image sensors consumes more energy than that on a backend processor. This is because the semiconductor fabrication technology, quantified by the \textit{process node}, of the sensor usually lags at least one generation behind that of the processor. Today, many commercial processors are fabricated using a 7~nm process node or smaller, but even high-end image sensors still use a 14~nm or 28~nm process node~\cite{kwon2020low, yu201914nm}. The computation power consumption increases \textit{quadratically} with respect to the process node~\cite{dennard1974design}. Therefore, one must carefully weigh the reduction of data transmission energy against the overhead of computing inside an image sensor on an older process node.

Our goal is thus to map the different components of our eye tracking algorithm to the sensor and processor in such a way that the total energy consumption is minimized. Intuitively, the ideal mapping is one where a trivial amount of computation in the sensor can drastically reduce the data communication.

To identify the optimal mapping, we first quantify the computation cost and the output data volume of the four main components in our algorithm: event map generation, edge map generation, prediction network, and segmentation network; the first three comprise the ROI prediction algorithm as shown in \Fig{fig:bbox_prediction}. \Tbl{table:pipeline_mac} shows the results. Note that each image pixel uses 1 byte; each pixel in the event map and the edge map uses 1 bit; and each segmentation map pixel uses 2 bits (4 classes).

Compared to the segmentation network, the entirety of the ROI prediction algorithm (columns 2--4 in \Tbl{table:pipeline_mac} combined) is much lighter weight, requiring less than 2\% of the Floating Point Operations (FLOPs). One straightforward solution would thus be to map the entire ROI prediction algorithm to the sensor while leaving only the segmentation network to execute on the processor chip. This mapping is shown in \Fig{fig:mode2}.

Compared to the mapping in today's systems, where all the computation takes place in the processor (illustrated in \Fig{fig:mode1}), predicting ROI in the sensor reduces the data transmission volume from the full-resolution image to only the ROI image (from sensor to processor) and the segmentation map (from the processor back to the sensor). The reduction in data transmission will outweigh the slight computation energy increase of moving the ROI prediction to the sensor, given the small computation requirements of the ROI prediction algorithm.

\begin{figure}[t]
    \centering
    \subfloat[No compute in sensor.]{
      \label{fig:mode1}
      \includegraphics[width=0.31\columnwidth]{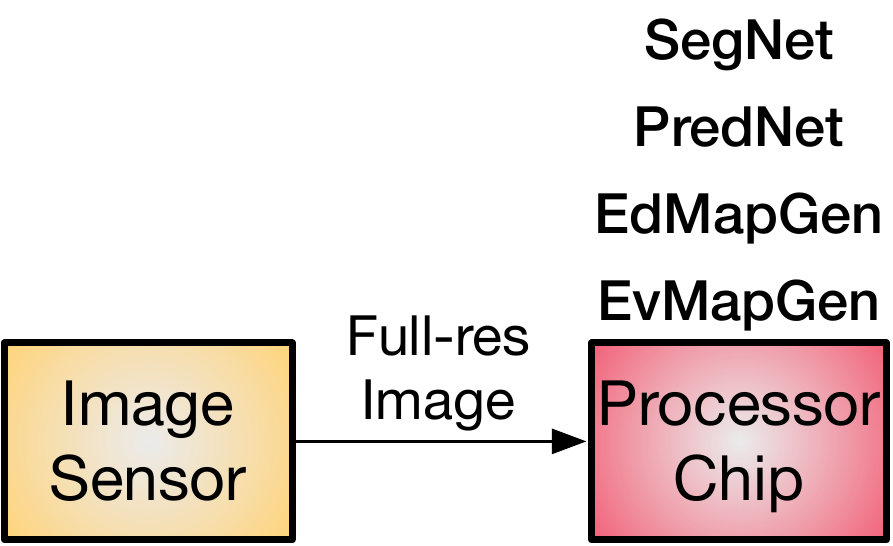}
    }
    \hfill
    \subfloat[Predict ROI in sensor.]{
      \label{fig:mode2}
      \includegraphics[width=0.31\columnwidth]{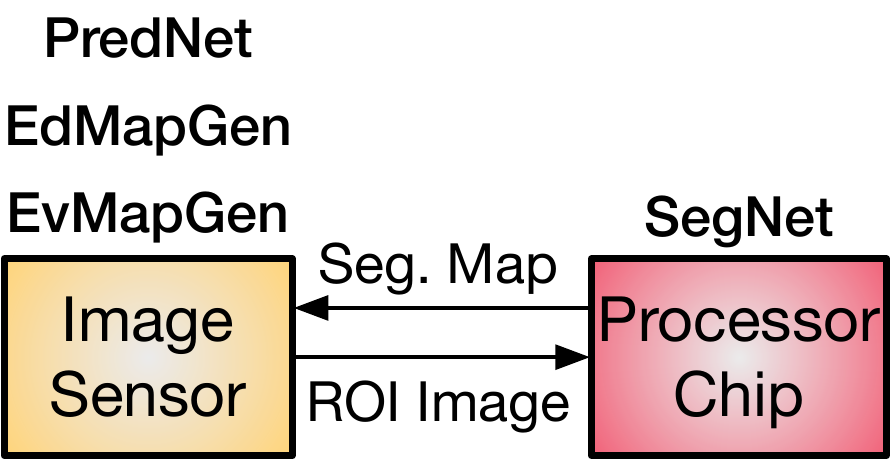}
    }
    \hfill
    \subfloat[Optimal mapping.]{
      \label{fig:mode3}
      \includegraphics[width=0.31\columnwidth]{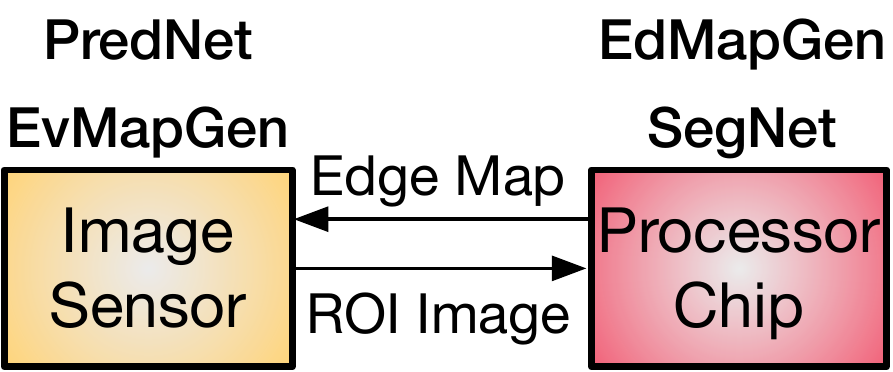}
    }
    \caption{Different hardware mapping schemes for the four algorithmic components: event map generation (EvMapGen), edge map generation (EdMapGen), ROI prediction network (PredNet), and eye segmentation network (SegNet). The optimal mapping (c) minimizes the total data transmission and computation energy.}
    \label{fig:mapping}
\end{figure}

Interestingly, mapping the ROI prediction network to the sensor is \textit{not} the most energy-efficient mapping. We propose a mapping, shown in \Fig{fig:mode3}, where only the event map generation and the prediction network execute in the sensor while the edge map generation and the segmentation execute on the processor. Compared to \Fig{fig:mode2}, our mapping reduces the data transmission volume (as the edge map, not the segmentation map, is transmitted) and the computation energy (as the edge map is generated in the processor).
}

%% file: roi_fig.tex
\begin{figure*}[t]
\centering
\includegraphics[width=\textwidth]{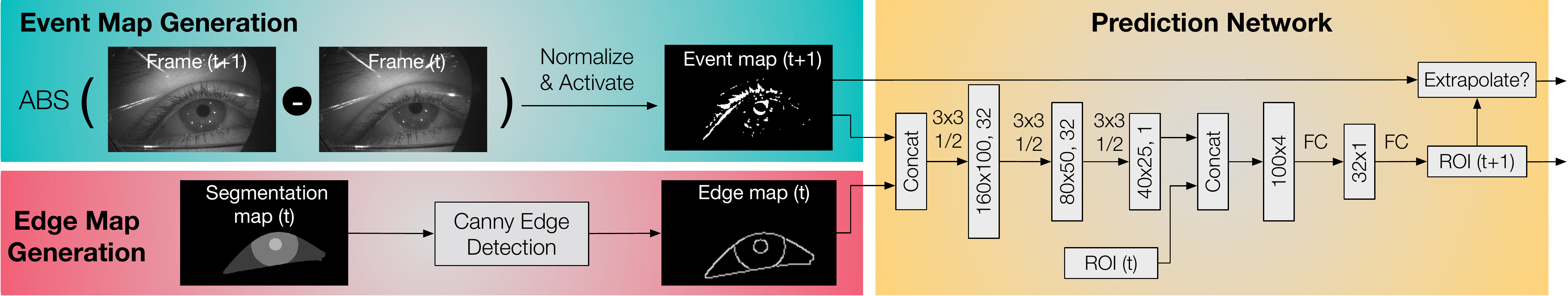}
\caption{The process of predicting the ROI at time $(T+1)$ consists of three steps. First, we compute the absolute difference of frames at $T$ and $(T+1)$ to generate an event map. Second, we use Canny edge detection on the segmentation map at time $T$ to extract an edge map. Finally, we concatenate the edge map and the event map to form the input to the ROI prediction network. The first three layers are Conv layers with $3\times3$ kernels followed by a Maxpool layer to reduce each dimension to $1/2$. The output of the last Conv layer is vectorized and concatenated with the ROI from time $T$ (a $1 \times 4$ vector). The concatenated vector then goes through two FC layers to generate the predicted ROI of time $(T+1)$. While our algorithm relies on events, the two additional cues of ROI and edge map from the previous frame are critical to the accuracy.}
\vspace{-5pt}
\label{fig:bbox_prediction}
\end{figure*}

%% file: seg_fig.tex
\begin{figure*}[t]
\centering
\includegraphics[width=1\textwidth]{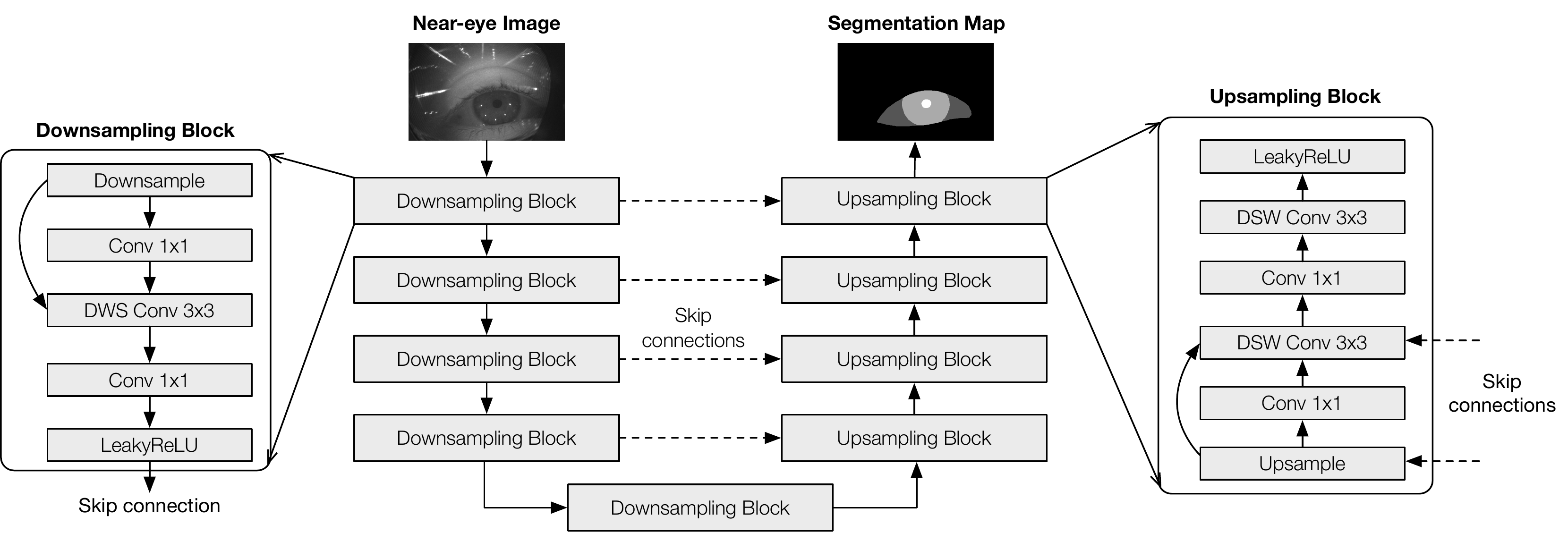}
\caption{The eye segmentation DNN. We use depthwise separable convolution as a building block to construct a U-Net-like architecture.}
\vspace{-10pt}
\label{fig:basic_blocks}
\end{figure*}

%% file: algorithm.tex
\subsection{Efficient Eye Segmentation Network}
\label{sec:framework:seg}

The Auto ROI mode avoids full-resolution segmentation almost altogether, but the segmentation stage is still the single biggest bottleneck in the system, contributing to over 85\% of the total execution time in our measurements. We propose a new segmentation network architecture that is much simpler without sacrificing accuracy.

\Fig{fig:basic_blocks} shows our segmentation network inspired by U-Net~\cite{ronneberger2015unet}. Our architecture contains five downsampling blocks and four upsampling blocks, similar to the original U-Net design. However, unlike the original U-Net and prior eye segmentation networks~\cite{chaudhary2019ritnet,yiu2019deepvog, kothari2021ellseg} that rely only on convolutional layers (\texttt{Conv}), we judiciously use depthwise separable convolution (\texttt{DWSConv})~\cite{chollet2017xception}, a lightweight convolution primitive used in many efficiency-oriented networks such as MobileNet~\cite{howard2017mobilenets} to balance compute efficiency with accuracy.
\no{
\begin{figure}[t]
\centering
\includegraphics[width=\columnwidth]{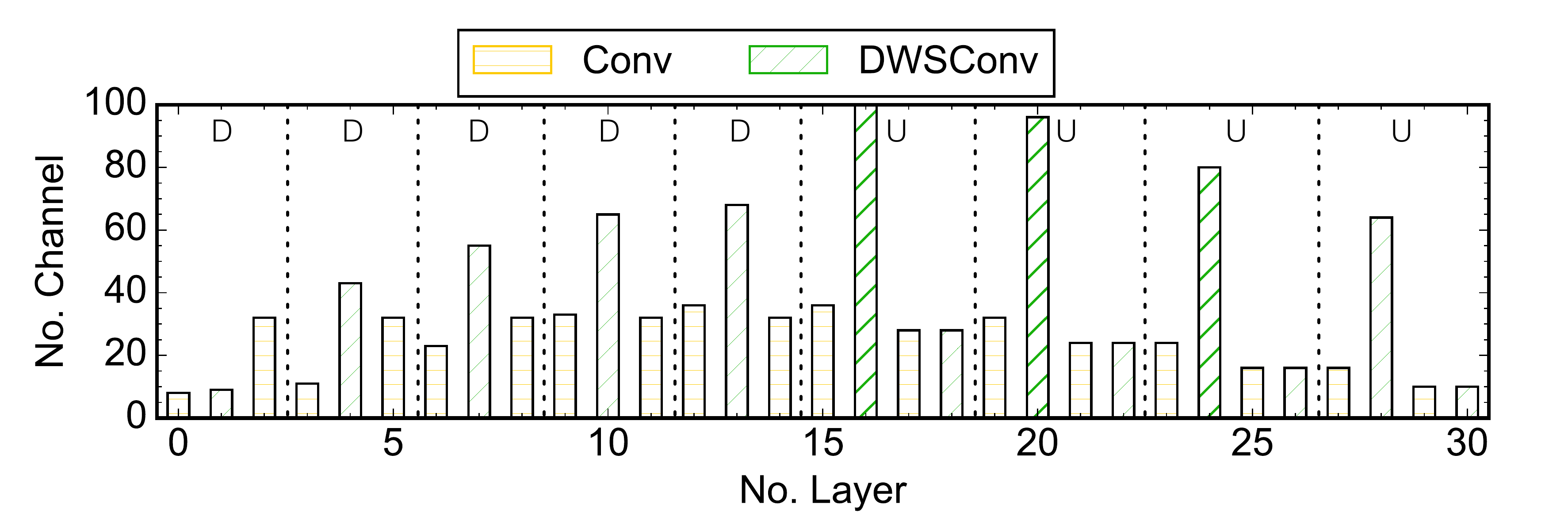}
\caption{Number of layer channels in each block in our eye segmentation network. The first five blocks are downsampling blocks (\texttt{D}) and the remaining four blocks are upsampling blocks (\texttt{U}). The \texttt{DWSConv} layers tend to have more channels than the others.}
\label{fig:pruned_cnt_each_layer}
\end{figure}
}

We interleave $1\times1$ \texttt{Conv} layers with $3\times3$ \texttt{DWSConv} layers. The \texttt{DWSConv} layers are used to filter different channels independently, and 1x1 \texttt{Conv} layers are used to combine different channels' features. This interleaving leverages the efficiency of \texttt{DWSConv} while allowing channels to learn across each other to ensure accuracy. This strategy is used in both downsampling and upsampling blocks. We use skip connections to facilitate gradient flow in our network.

The convolution channel width, both in the \texttt{Conv} and the \texttt{DWSConv} layers, affects the network's ability to learn. We empirically find that \texttt{DWSConv} layers with skip connections usually require a higher channel width compared to other layers, because the learning abilities of those layers have global effects. \no{This design decision is shown in \Fig{fig:pruned_cnt_each_layer}, which shows the layer-wise configuration of our network.} As we will show, the channel width is an effective knob that provides a speed-vs-accuracy trade-off.



%% file: setup.tex
\section{Evaluation Methodology}
\label{sec:exp}

\input{dataset}

\input{res_fig}

\subsection{Training}
\label{sec:exp:training}

We first train a standalone eye segmentation network and a standalone ROI prediction network, and then refine eye segmentation using the ROI prediction results.

We follow prior studies and split the entire dataset into the training/validation/test sets with a 80/20 training/validation split, identical to EllSeg~\cite{kothari2021ellseg}. The eye segmentation network is trained using the Adam optimizer, a learning rate of 0.001, and a batch size of 4 for 250 epochs. We use the a loss function that combines both the standard cross-entropy loss and losses that are specialized to the eye structure~\cite{chaudhary2019ritnet}. To train the ROI prediction network, we use the Adam optimizer with a learning rate of 0.001, a momentum of 0.9, and the mean squared error loss function. We train the ROI prediction network for 100 epochs with a batch size of 8.

We then fine-tune the eye segmentation network using ROI images. In this step, the ROI prediction network first predicts an ROI given an eye image; we then crop the input image based on the predicted ROI and fine-tune the segmentation network with the cropped image. This training procedure is similar to networks based on region proposals such as Faster R-CNN~\cite{ren2015faster}. We use a learning rate of 0.0001, a momentum of 0.9, and 100 epochs.

\subsection{Baseline and Evaluation Metrics}
\label{sec:exp:baseline}

We primarily compare against eye segmentation methods that are based on DNNs, which are shown to have superior accuracy compared to non-DNN-based segmentation methods~\cite{akinlar2021accurate, fuhl2016pupilnet}. The baselines are trained using the same procedure as our algorithm.
\begin{itemize}
  \setlength \itemsep{-3pt}
  \item \texttt{RITnet}~\cite{chaudhary2019ritnet}: an encoder-decoder network using DenseNet \cite{huang2017densenet} as the backbone. It won first place in the 2019 OpenEDS segmentation challenge~\cite{openeds2019}. As discussed in \Sect{sec:exp:dataset}, we use RITnet to generate the ground truth for the sequential OpenEDS data. As a result, it is expected that our algorithm will have lower accuracy than \texttt{RITnet}.
  \item \texttt{DenseElNet}~\cite{kothari2021ellseg}: an ellipse segmentation framework for gaze tracking. \texttt{DenseElNet} is originally for ellipse segmentation (three segments). We re-purpose \texttt{DenseElNet} to predict four segments by modifying the channel size of the last layer.
  \item \texttt{DeepVOG}~\cite{yiu2019deepvog}: a popular encoder-decoder network for eye segmentation. \texttt{DeepVOG} was originally designed to generate only two segments (pupil and background). We modified the channel size of the last layer so that it predicts four segments.
\end{itemize}


We also compare the speed of our algorithms with the baselines. We measure the speed on a state-of-the-art mobile computing platform, Nvidia's Jetson Xavier board~\cite{xavier2dev}. We use the mobile Volta GPU on the Xavier board for all the networks. The GPU has 512 CUDA cores with a maximum frequency of 1,377 MHz.

\subsection{Variants}
\label{sec:exp:variants}

We use two DNN variants of our eye segmentation network, \texttt{Ours(S)} and \texttt{Ours(L)}. Both are designed with the same architecture, but differ in the channel width and thus provide a speed-vs-accuracy trade-off. In particular, \texttt{Ours(S)} has, on average, half the channels of \texttt{Ours(L)}. \Tbl{table:model_size} compares the amount of Floating Point Operations (FLOPs) and parameters of the two networks against the baseline segmentation networks.

\begin{table}[ht]
\caption{FLOPs and number of parameters in different eye segmentation networks for an input size of $640\times400$ 8-bit pixels.}
\centering
\resizebox{\columnwidth}{!}{
\renewcommand*{\arraystretch}{1}
\renewcommand*{\tabcolsep}{2pt}
\begin{tabular}{c c c c c c}
\toprule[0.15em]
\textbf{Network} & \textbf{DenseElNet} & \textbf{DeepVOG} & \textbf{RITnet}  & \textbf{Ours (L)} & \textbf{Ours (S)} \\
\midrule[0.05em]
FLOPs (Billion) & 53.1 & 36.5 & 23.1 & 2.6 & 1.2 \\
Norm. FLOPs & 45.2 & 31.1 & 19.7 & 2.3 & 1.0 \\
\midrule[0.05em]
\# of Parameters (Thousand) & 3047.3 & 2835.7 & 391.0 & 73.0 & 30.6 \\
Norm. \# of Parameters & 99.6 & 92.6 & 12.8 & 2.4 & 1.0 \\
\bottomrule[0.15em]
\end{tabular}
}
\label{table:model_size}
\end{table}

The total computation requirement of \texttt{Ours(L)} is only about 1/9 of \texttt{RITnet}, 1/14 of \texttt{DeepVOG}, and 1/20 of \texttt{DenseElNet}. \texttt{Ours(S)} is even lighter weight, with about 45$\times$ fewer FLOPs and 100$\times$ fewer parameters compared to \texttt{DenseElNet}. Overall, \texttt{Ours(S)} uses only about 30K parameters ($\sim$120 KB).

To tease apart the contributions of the different components of our algorithm, we evaluate two variants on top of eye segmentation:
\begin{itemize}
  \setlength \itemsep{-3pt}
  \item \texttt{+ROI}: this variant uses only the ROI prediction (\Sect{sec:framework:roi:pred})
  \item \texttt{+ROI+E}: this variant uses both ROI prediction (\Sect{sec:framework:roi:pred}) and extrapolation  (\Sect{sec:framework:roi:extra})
\end{itemize}

%% file: dataset.tex
\subsection{Dataset and Ground Truth}
\label{sec:exp:dataset}

\paragraph{OpenEDS 2020}
We use the sequential data from the OpenEDS 2020~\cite{palmero2020openeds2020} dataset (Track-2)\footnote{We do not use Track-1 data as the videos are only 1--2 seconds long.}, which contains 200 sequences of continuous frames (29,476 frames in total) gathered from 152 participants with various ethnicities and iris colors. All sequences are captured by an infrared camera at 100 Hz with a 640$\times$400 resolution. The average duration of a sequence is 30 seconds with diverse eye movements such as blinks and saccades (rapid eye movements). The average number of blinks per video is 4.5 (up to 18) and the average number of saccades per video is 7.8.

\no{
\begin{figure}[t]
\centering
\includegraphics[width=\columnwidth]{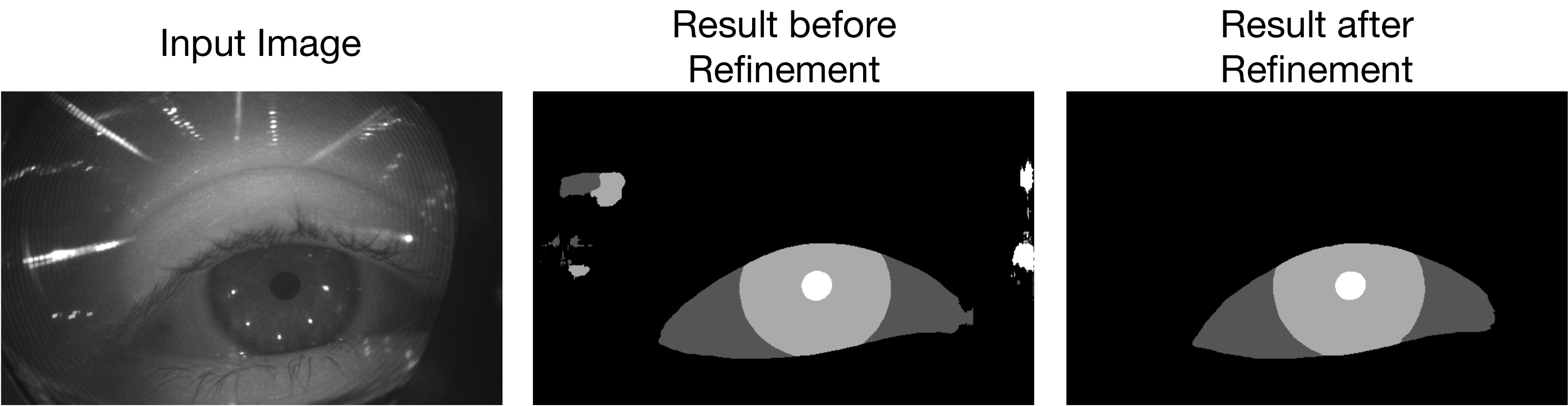}
\caption{Ground truth generation. We use RITnet~\cite{chaudhary2019ritnet} to generate an initial segmentation (middle) from the input eye image (left). The initial result is noisy, which we refine using a spatial clustering algorithm~\cite{ester1996density} to obtain the final segmentation ground truth (right).}
\label{fig:refine}
\end{figure}
}

However, the sequential data in OpenEDS 2020 do not include the segmentation ground truth. To generate the ground truth for our evaluation, we train RITnet~\cite{chaudhary2019ritnet}, a state-of-the-art eye segmentation network, on the non-sequential OpenEDS 2019~\cite{garbin2019openeds} dataset (100 Hz with a 400$\times$640-pixel resolution), which does have ground truth segmentation labels. We then use the trained RITnet to generate the eye segmentation results for the sequential OpenEDS 2020 data.

The generated eye segmentation results are not perfect; we perform a series of data refinement steps to generate the final ground truth. We first apply the DBSCAN spatial clustering algorithm~\cite{ester1996density} on each initial segmentation result and identify the largest contiguous region as the correct eye region. We then fill the missing holes inside the eye region as a prior study does~\cite{kim2019eye}. Finally, we manually inspected all sequences of the refined data and removed 15 sequences (out of 200), where the eye segmentations are visibly incorrect. In the end, our sequential dataset contains 185 sequences and 27,431 total frames.
\no{\Fig{fig:refine} shows an example before and after refinement, where the disconnected regions in the initial result are removed, providing a more faithful ground truth.}

With the eye segmentation ground truth, we then apply a gaze estimation algorithm~\cite{Swirski2013} commonly used in prior work (e.g., DeepVOG~\cite{yiu2019deepvog}) to generate the gaze ground truth. Our algorithm does not depend on a particular gaze estimation method and, thus, can be integrated with other gaze estimation models~\cite{dierkes2019fast}. We also use the eye segmentation ground truth to generate the ROI ground truth, which is the bounding box of the foreground segment (pupil, iris, and sclera). We manually verify that using so-defined ground-truth ROIs gives the same gaze results as those from full-frame inference.

\paragraph{TEyeD} Another dataset we used in our evaluation is the TEyeD dataset, which has manually annotated segmentation and gaze ground truth~\cite{fuhl2021teyed}. We select 2 sequences from the dataset. Both sequences are grayscale images at a 384$\times$288 resolution of captured at 25 Hz with an average length of over 30 minutes. In total, the selected data provide over 100,000 images.

On the TEyeD dataset, we report the 2D pupil position error (in pixels) rather than the 3D gaze direction error. This is because the gaze estimation algorithm~\cite{Swirski2013} we use, not a contribution of this paper, does not provide accurate gazes for the TEyeD dataset.

%% file: res_fig.tex
\begin{figure*}[t]
  \centering
    \subfloat[\small{Vertical gaze error vs. speedup.}]{
    \label{fig:vgaze_speedup}
    \includegraphics[height=1.0in]{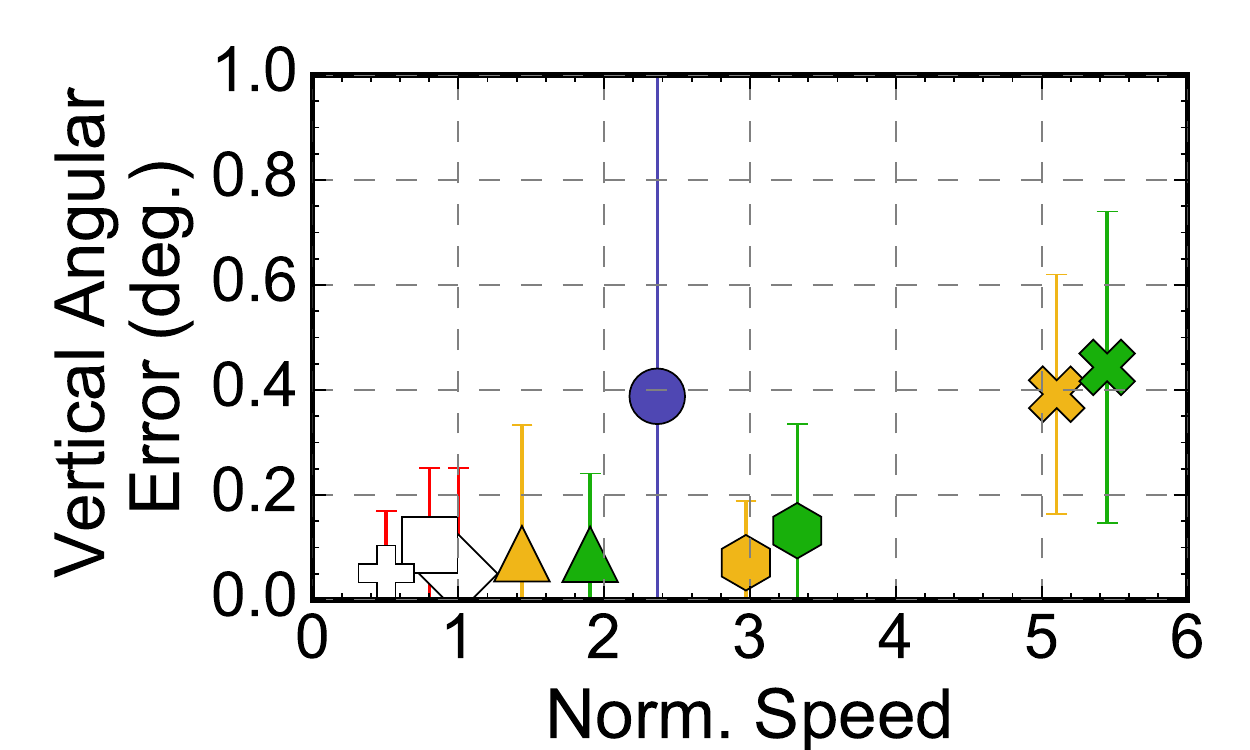} }
    \subfloat[\small{Horizontal gaze error vs. speedup.}]{
	\label{fig:hgaze_speedup}
	\includegraphics[height=1.0in]{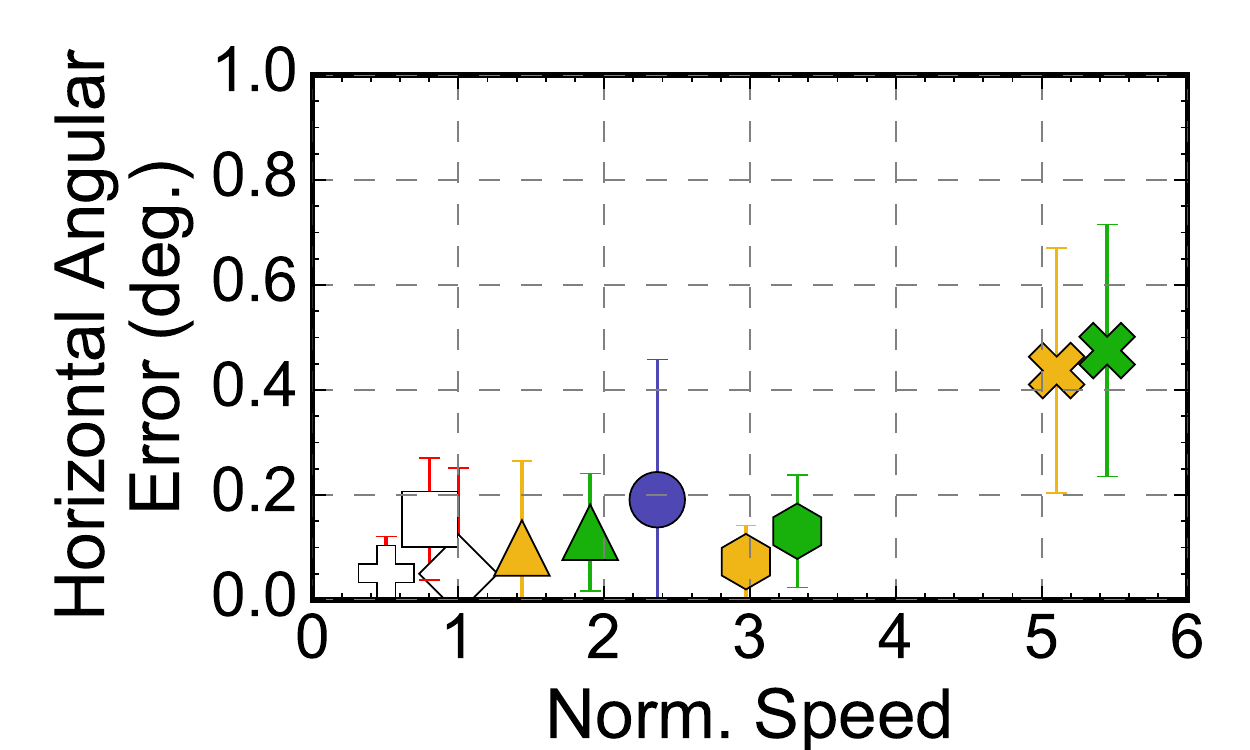}} 
    \subfloat[\small{mIoU vs. speedup.}]{
    \label{fig:miou_speedup}
    \includegraphics[height=1.0in]{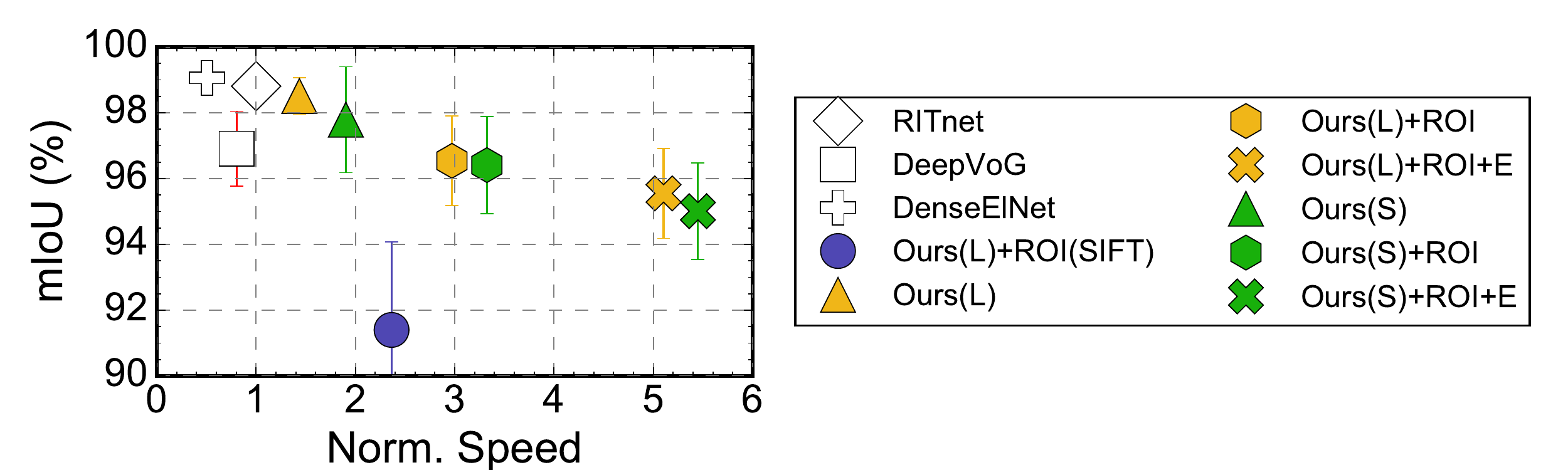}}
    \caption{The accuracy and speed comparison of different methods. All the subfigures share the same legend. The speedup values are normalized to the speed of \texttt{RITnet}. \texttt{Ours(S)} and \texttt{Ours(L)} are two eye segmentation networks. \texttt{+ROI} denotes ROI prediction is enabled. \texttt{+E} denotes the extrapolation is enabled. \texttt{+ROI(SIFT)} denotes using the SIFT-based ROI prediction.}
    \label{fig:gaze_speedup}
\end{figure*}

%% file: eval.tex
\section{Evaluation}
\label{sec:eval}

We first use OpenEDS 2020 to drive the analysis of the results. We present the overall performance and accuracy comparison (\Sect{sec:eval:perf}), followed by analyzing the sensitivity of our system to various parameters (\Sect{sec:eval:sen}) and algorithmic choices (\Sect{sec:eval:design}). We perform an ablation study to evaluate the importance of the two feedback cues (\Sect{sec:eval:ablation}). We present the results from the TEyeD dataset (\Sect{sec:eval:teyed}) and analyze failure cases (\Sect{sec:eval:fail}). Finally, we present the results of in-sensor auto-ROI (\Sect{sec:eval:in_sensor}).

\subsection{Accuracy vs. Speed}
\label{sec:eval:perf}

\begin{figure}[t]
\centering
\includegraphics[width=\columnwidth]{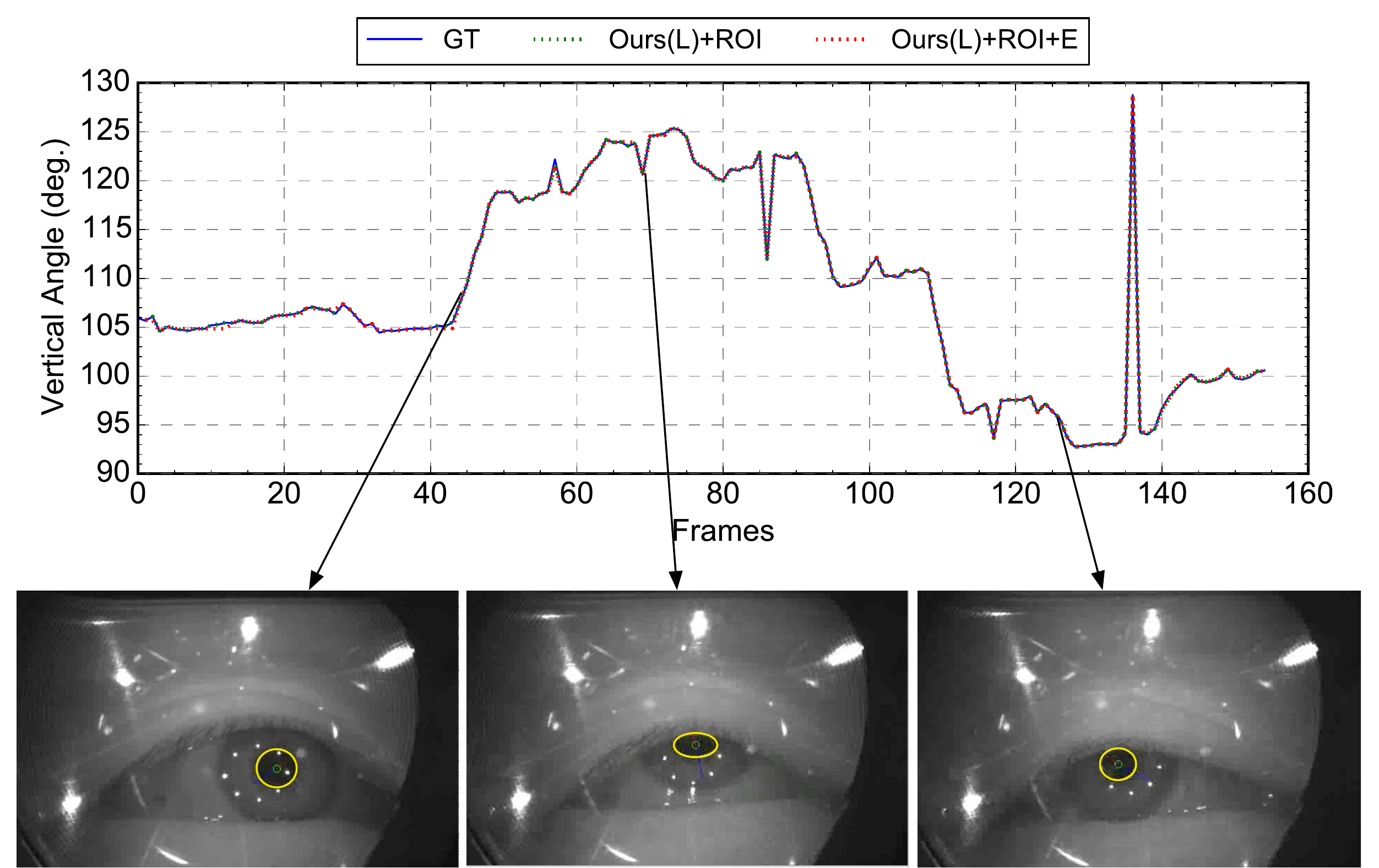}
\caption{Gaze estimation results over one sequence of frames. \texttt{Ours(L)} robustly tracks the ground truth. The bottom panel shows three representative cases: eye moves right, just before a blink, and eye moves up-left, respectively.}
\label{fig:angle_sequence}
\end{figure}

\paragraph{Gaze Estimation} Our algorithm achieves a 5.5$\times$ speedup over the baselines with a sub-\ang{0.5} gaze error. \Fig{fig:vgaze_speedup} and \Fig{fig:hgaze_speedup} compare the vertical and horizontal gaze errors and the speed of our algorithms with different baselines. The speedups are normalized to the speed of \texttt{RITnet}, which runs at 5.4 Hz on a mobile Volta GPU. Note that a \ang{1} error is generally acceptable for gaze tracking~\cite{kar2017review}.

\texttt{RITnet}, \texttt{DenseElNet}, \texttt{Ours(L)}, and \texttt{Ours(S)} keep the absolute error rate below \ang{0.1} in both the vertical and horizontal direction. They are all more accurate than \texttt{DeepVOG}. In particular, \texttt{Ours(S)} only has 0.04 and 0.05 higher gaze error than \texttt{RITnet} in each direction, but is 1.9$\times$ faster. Note that this little gaze error loss is achieved with a network that is 12.8$\times$ smaller (\Tbl{table:model_size}). This highlights our efficient eye segmentation network design (\Sect{sec:framework:seg}).

By operating on ROI images when possible, \texttt{Ours(L)+ROI} improves the speedup over \texttt{RITnet} to 3.0$\times$. Interestingly, \texttt{Ours(L)+ROI} achieves better accuracy than \texttt{Ours(L)}. Further inspection of the data shows that this is because by using only the ROI image for eye segmentation, we remove potential noise in the non-ROI region in the input eye image.

Using activity-proportional segmentation, \texttt{Ours(L)+ROI+E} and \texttt{Ours(S)+ROI+E} further improve the speedup to $4.2\times$ and $5.5\times$, respectively, while both retaining an angular error rate within \ang{0.5}. The event density threshold used for extrapolation is set to 0.1\%. We will study the algorithm's sensitivity to this parameter in \Sect{sec:eval:sen}.

To further confirm the robustness of our system, \Fig{fig:angle_sequence} compares the frame-by-frame gaze results of \texttt{Ours(L)}, \texttt{Ours(L)+ROI+E}, and the ground truth. \texttt{Ours(L)} virtually matches the ground truth, and \texttt{Ours(L)+ROI+E} has slight deviations (e.g., around frame 10). We show three representative gazes in the bottom panel, where the eye moves moves right, blinks, and moves up left.

\begin{figure}[t]
\centering
\includegraphics[width=\columnwidth]{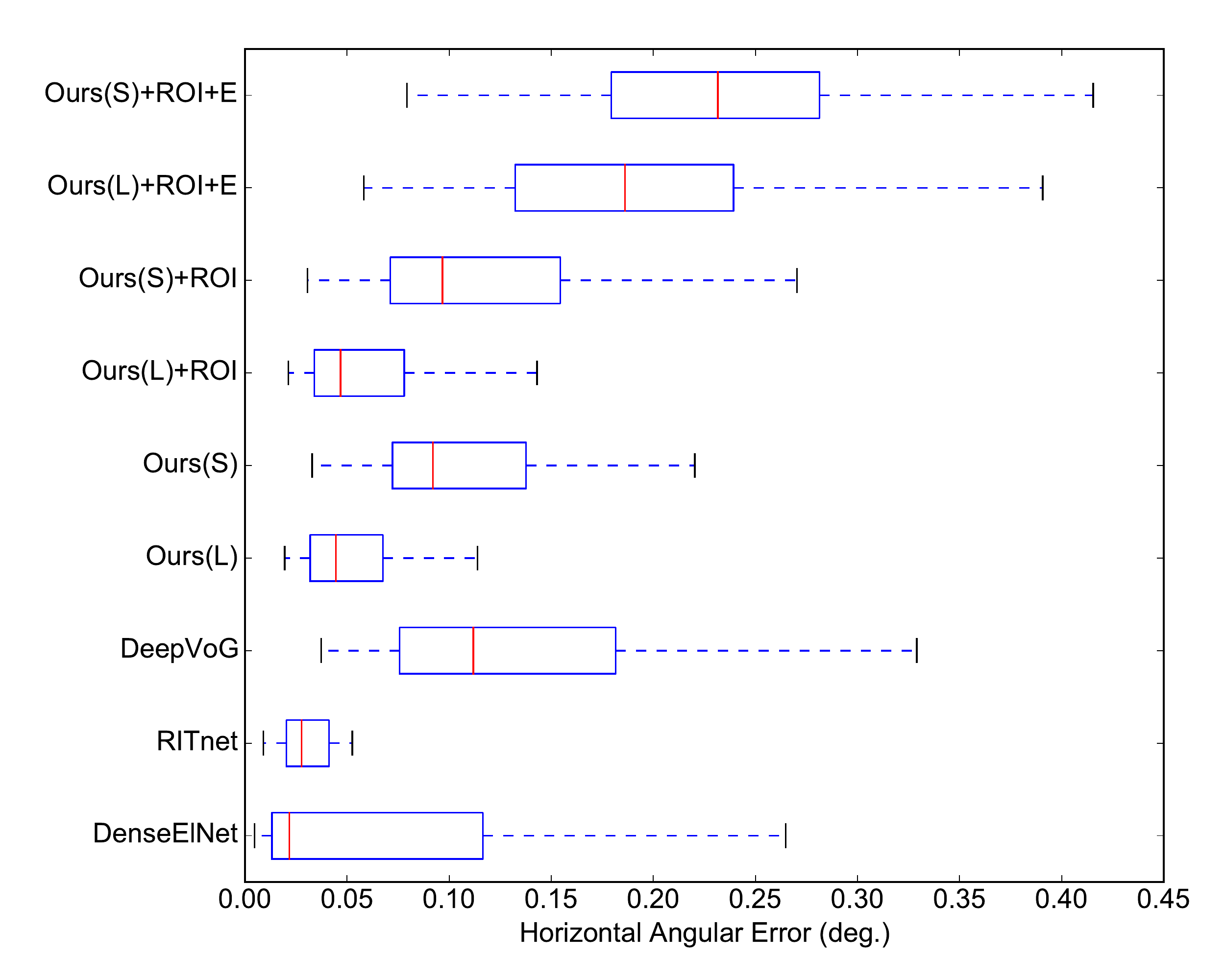}
\caption{Distributions of horizontal gaze error across as boxplots, which plot the median, 25th-percentile, 75th-percentile, the min, and max of the angular errors. \texttt{RITNet} is the most accurate because it is used to obtain the ground truth (see \Sect{sec:exp:dataset}). Even the most inaccurate variant of our system, \texttt{Ours(S)+ROI+E} has a worst-case accuracy below \ang{0.5}, which is generally regarded as an acceptable error bound for eye tracking~\cite{kar2017review}.}
\label{fig:gazedist}
\end{figure}

Finally, for a comprehensive analysis we show the gaze error distributions across all the evaluated frames in \Fig{fig:gazedist}. Not surprisingly, \texttt{RITNet} has the most compact distribution, because it is used to obtain the ground truth (see \Sect{sec:exp:dataset}).  \texttt{Ours(L)} has significantly better accuracy distribution compared to \texttt{DeepVOG}. Comparing to \texttt{DenseElNet}, \texttt{Ours(L)} has \ang{0.02} lower average accuracy, but also has much better worst-case accuracy, as indicated by the significantly shorter tail. As we use a smaller network and exploit more speed-enhancing techniques, the error distribution generally shifts toward the right, but even the most inaccurate variant, \texttt{Ours(S)+ROI+E}, has a worst-case accuracy lower than \ang{0.5}, which is regarded as an acceptable error bound for eye tracking~\cite{kar2017review}.

\paragraph{Eye Segmentation} While gaze accuracy is ultimately what we care about, we show the results of eye segmentation, an intermediate step to gaze estimation, to understand how the two metrics relate.

\Fig{fig:miou_speedup} compares the eye segmentation accuracy and speed across different methods. 
\texttt{DenseElNet} achieves the highest mIoU at $99.0\%$. \texttt{Ours(L)} achieves a similar mIoU at $98.6\%$, but is 2.8$\times$ faster and 20$\times$ smaller. With a smaller network configuration, \texttt{Ours(S)} introduces a $1.2\%$ mIoU loss. \texttt{Ours(L)+ROI} and \texttt{Ours(S)+ROI} have slightly lower mIoU at $96.5\%$ and $96.4\%$, respectively. With extrapolation enabled, \texttt{Ours(L)+ROI+E} and \texttt{Ours(S)+ROI+E} can still keep the mIoU loss within $1.5\%$. Overall, our algorithm has slightly higher accuracy loss over \texttt{RITnet} on the segmentation metric than on the gaze error metric. This shows that the gaze estimation, a model fitting problem, can tolerate segmentation inaccuracy.

\paragraph{Frame Distribution} On \texttt{Ours(L)+ROI+E}, 44.9\% of frames perform extrapolations, 54.5\% of frames are processed in the ROI mode, and only 0.6\% of frames require full-resolution eye segmentation. The vast majority of the frames that require full-resolution processing are the initial frames in each sequence. This frame distribution explains the speedup of our algorithm.

\paragraph{Time Distribution} The time distribution of different stages of our algorithm points out opportunities for future optimizations. When executing \texttt{Ours(L)+ROI+E} on the mobile Volta GPU, the most time-consuming stage is the segmentation network, which takes 66.9\% of total time. The second most time-consuming stage is the ROI prediction network, which takes 18.7\% of the time. The event map and edge map are calculated on the CPU and take 3.6\% and 3.9\% of the time, respectively. The rest of the algorithm (data marshalling) takes 6.1\% of the time.

We emphasize that speed we report in the paper should be seen as a \textit{lower bound} of our algorithm and can be improved with more engineering efforts. For instance, the ROI prediction and eye segmentation networks can be pipelined across frames, and redundant memory copies can be removed if we directly manage the GPU memory rather than through PyTorch. Our goal here is to show that even unoptimized code (speed-wise) achieves several-fold speedup over state-of-the-art algorithms with virtually the same gaze accuracy.

\input{sen_fig}
\subsection{Sensitivity Study}
\label{sec:eval:sen}

\paragraph{Event Activation Threshold} Our algorithm emulates how an event camera generates events. The event activation threshold ($\sigma$ in \Eqn{eqn:event_func}) dictates how sensitive our algorithm is to eye activities. So far our evaluation has used a $\sigma$ of $30\%$ as the default value.

\Fig{fig:event_threshold_sensitivity} shows how $\sigma$ affects the speed-vs-gaze trade-off, as $\sigma$ increases from 15\% to 90\%. We show only the horizontal error; results for vertical angular error are similar and have been omitted. We retrain the ROI prediction network for each $\sigma$ for a fair comparison. The speed is normalized to that of \texttt{RITnet}. The gaze errors under all $\sigma$ settings are consistently lower than \ang{0.5}, indicating an overall robust algorithm. 30\% is an empirical sweep spot. A lower threshold, e.g., 15\%, degrades accuracy, because a small threshold generates a noisy event map. Conversely, a higher threshold, e.g., 90\%, has lower accuracy too, because a large threshold generates a sparse event map.

\paragraph{Event Density Threshold} \Fig{fig:extrapolation_threshold_senstivity} shows how the horizontal gaze error and speed vary with the event density threshold ($\gamma$), which we use to determine when to extrapolate (\Sect{sec:framework:roi:extra}). $\gamma$ is increased from 0.001\% to 0.1\% from left to right. As $\gamma$ increases, more frames are extrapolated, so both the speed and the error increase. The gaze errors are consistently below \ang{0.5}. Under a 0.001\% $\gamma$, we achieve $3.9\times$ speedup over \texttt{RITnet}.

\begin{figure}[t]
\centering
\includegraphics[width=\columnwidth]{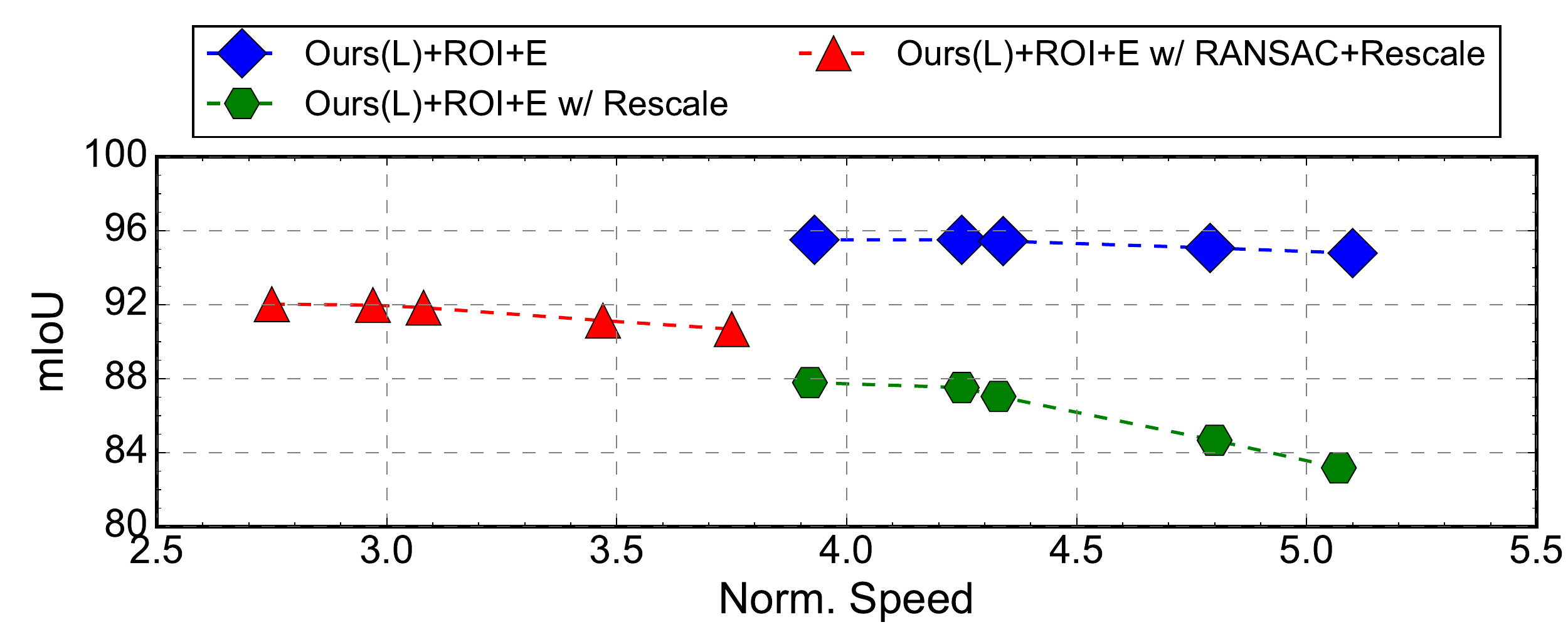}
\caption{Impact of extrapolation method on segmentation mIoU and speed normalized to \texttt{RITnet}. Extrapolation threshold values from left to right are 0.001\%, 0.005\%,0.01\%, 0.05\%, 0.1\%.}
\label{fig:extrapolation}
\end{figure}

\subsection{Design Decision Analysis}
\label{sec:eval:design}

\paragraph{ROI Prediction Method} To understand the effectiveness of our ROI prediction network, we construct a non-DNN method similar to those used in classic image stitching algorithms~\cite{brown2007automatic}. This method predicts the current frame's ROI by matching SIFT features~\cite{lowe1999object} from two consecutive image frames and using RANSAC~\cite{fischler1981random} to calculate a transformation matrix between the two frames. Using the transformation matrix, we then compute the ROI coordinates of the current frame from the previous ROI. This variant is denoted \texttt{Ours(L)+ROI(SIFT)}.


The gaze accuracy and the speed of \texttt{Ours(L)+ROI(SIFT)} are shown in \Fig{fig:gaze_speedup} in comparison to other methods. The gaze error of \texttt{Ours(L)+ROI(SIFT)} increases significantly, especially on the vertical direction, to 0.39, much higher than \texttt{Ours(L)+ROI} (0.08). This suggests the effectiveness of our ROI prediction DNN.

\paragraph{Extrapolation Method} The extrapolation method that we use in \Sect{sec:framework:roi:extra} simply reuses the previous segmentation map. To demonstrate its effectiveness, we compare with two other methods:

\begin{itemize}
  \setlength \itemsep{-3pt}
  \item \texttt{Rescale}: We first crop the previous segmentation map based on the previous ROI, and rescale the cropped segmentation map based on the size of the current, predicted ROI.
  \item \texttt{RANSAC+Rescale}: We use the ROI prediction network to first predict the ROI. The ROI is refined based on the SIFT features using RANSAC from the previous ROI. Lastly, we perform the same \texttt{Rescale} method as above.
\end{itemize}

\Fig{fig:extrapolation} compares the speedup (over \texttt{RITnet}) and segmentation accuracy of different extrapolation methods. All the extrapolation methods are applied to the \texttt{Ours(L)} network. We also vary the extrapolation threshold from 0.001\% (left-most markers, fewest extrapolations) to 0.1\% (right-most markers, most extrapolations).

We make two observations. First, our extrapolation method is consistently the most accurate. Combining RANSAC with rescaling is more accurate than simple rescaling, but also is much slower due to the overhead of RANSAC. Second, the accuracy generally drops with more extrapolations. But our method is the most robust to extrapolation. With 37.5\% more extrapolations the mIoU degrades by only 0.7\% (with 1.3$\times$ speedup), whereas the accuracy under rescaling drops by 4.6\%. The accuracy drop in rescaling-based methods is likely due to ROI error accumulation.

\input{acc_table}

\subsection{Ablation Study}
\label{sec:eval:ablation}

\begin{figure}[t]
  \vspace{-10pt}
  \centering
  \subfloat[Impact of previous ROI.]{
    \label{fig:miou_bbox_sensitivity}
    \includegraphics[width=0.48\columnwidth]{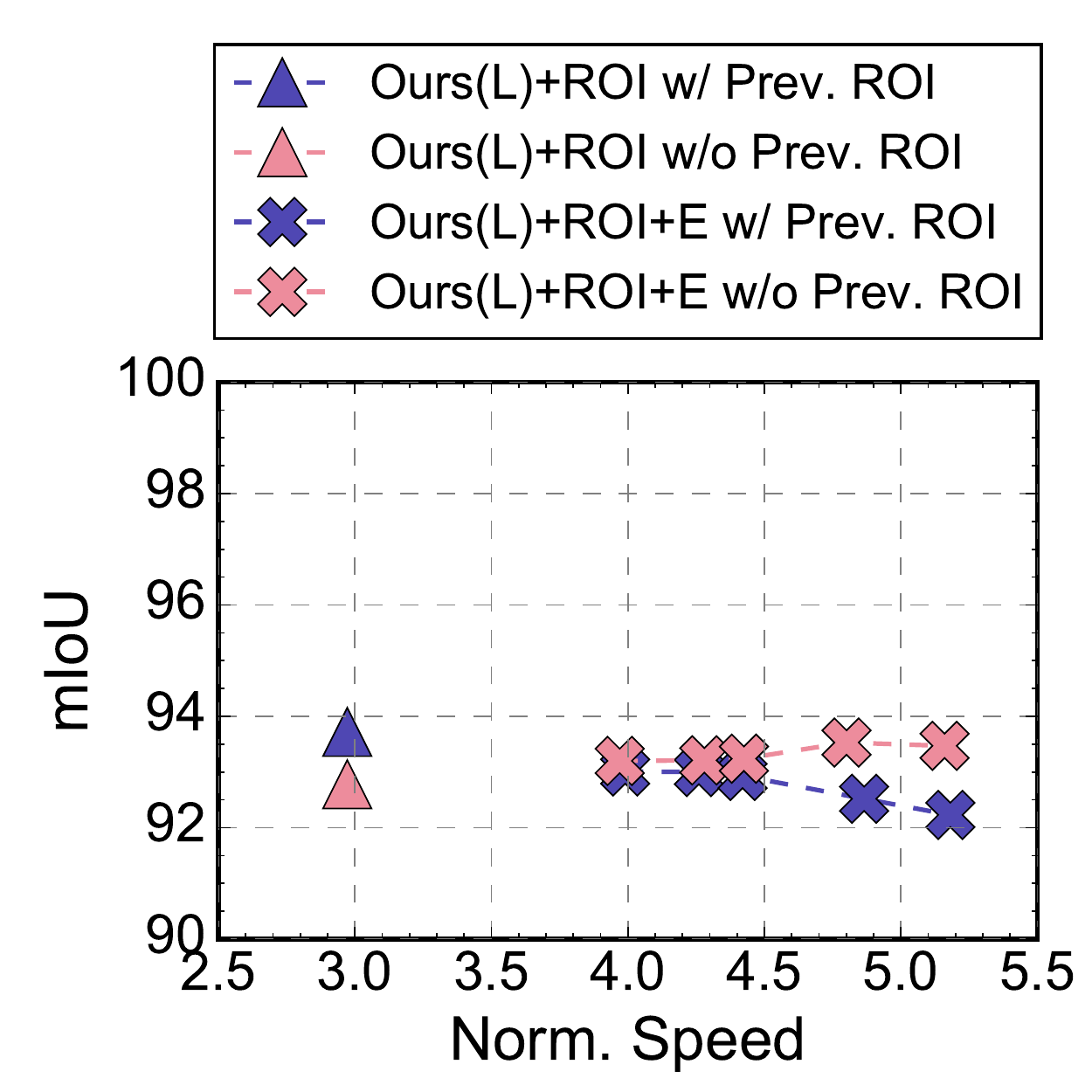}
  }
  \hfill
  \subfloat[Impact of edge map.]{
    \label{fig:miou_edge_sensitivity}
    \includegraphics[width=0.48\columnwidth]{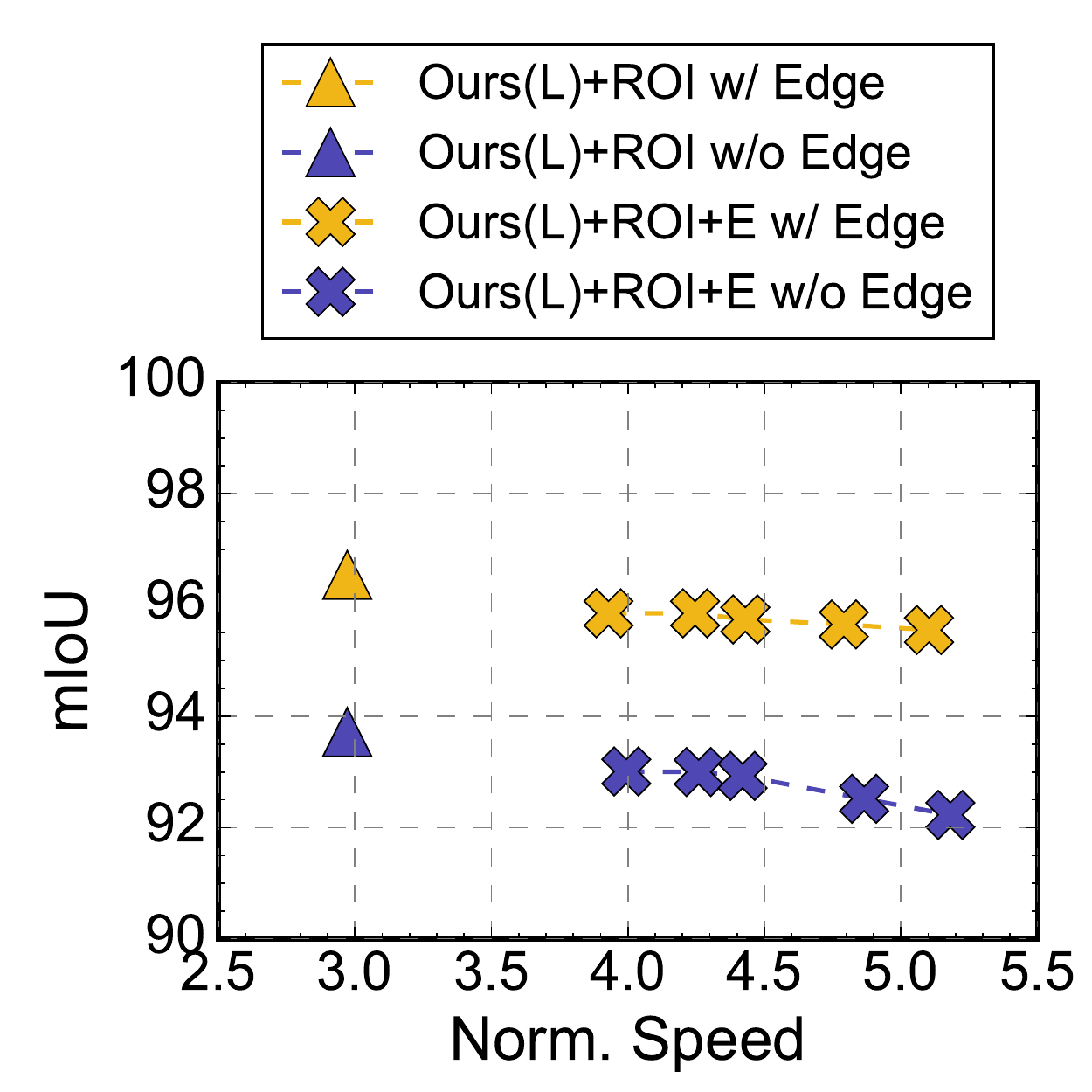}
  }
  \caption{Impact of using previous ROI and edge map in mIoU. Speeds are normalized to that of \texttt{RITnet}.}
  \label{fig:sen}
\end{figure}

\begin{figure}[t]
\centering
\includegraphics[width=.95\columnwidth]{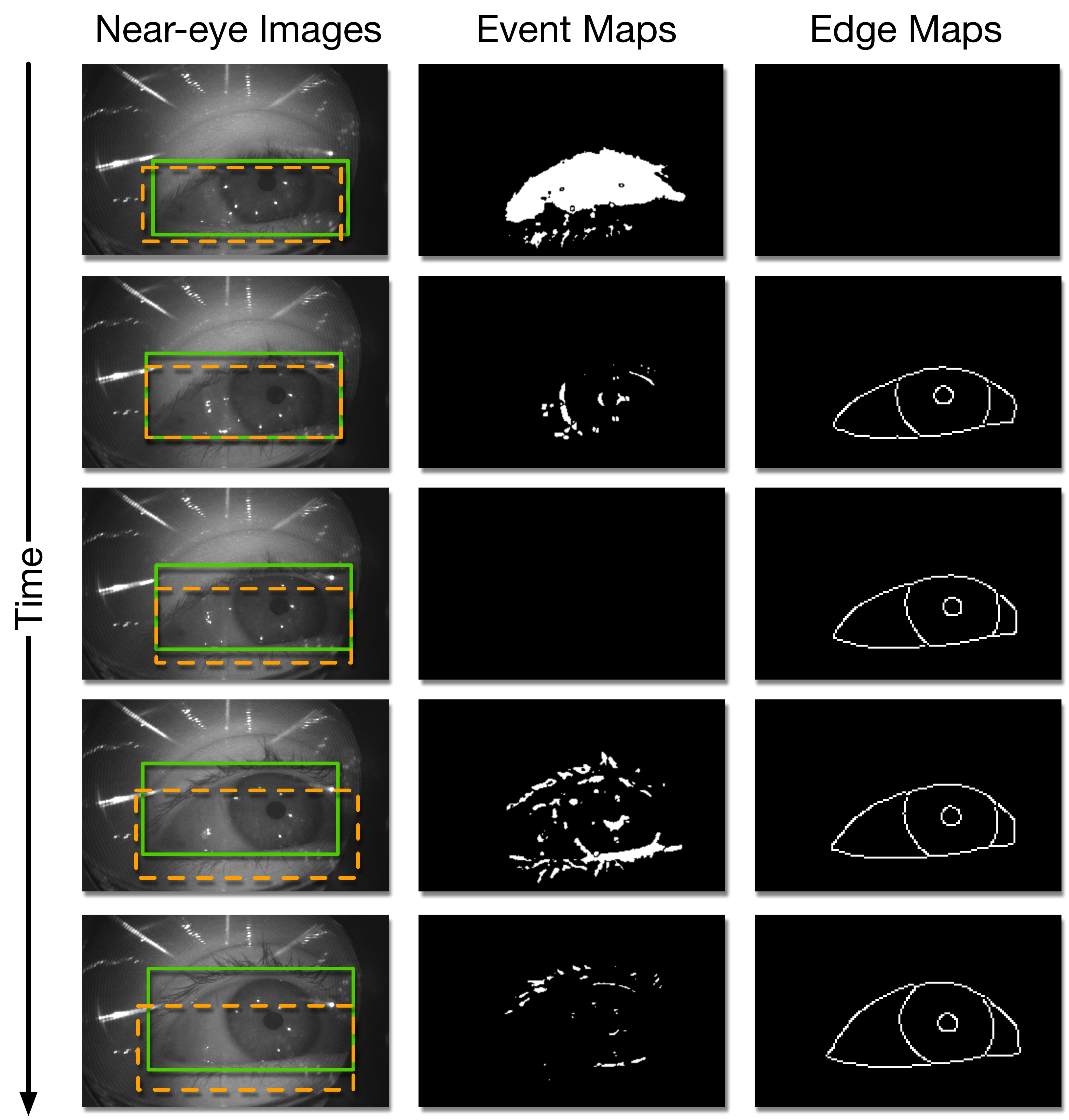}
\caption{A sequence showing that using the edge map helps correcting drifted ROIs. The solid and dashed rectangles are the ROIs predicted with and without using the edge map.}
\label{fig:seq_examples}
\end{figure}

The ROI prediction algorithm uses not only the event map, but also two feedback cues from the previous frame, i.e., the edge map and the ROI. Ablation studies here show the need for these two cues.

\paragraph{Previous ROI} \Fig{fig:miou_bbox_sensitivity} compares two ROI prediction networks: one only uses event maps and one uses event maps and previous ROIs. We train two ROI prediction networks using the same training procedure. We compare the mIoU and the speed of these two prediction schemes both with (\texttt{+ROI+E}) and without (\texttt{+ROI}) extrapolation. In the extrapolation mode, we sweep the event density threshold $\gamma$ from 0.001\% through 0.1\%, same as in \Fig{fig:extrapolation_threshold_senstivity}. Without extrapolation, using the previous ROI improves the segmentation mIoU by about 1\%, confirming that ROI feedbacks help predict future ROIs.

With extrapolation, the mIoU from using the previous ROI drops as $\gamma$ increases. This shows how ROI prediction errors can accumulate. Interestingly, the mIoU without using the previous ROI \textit{increases} as $\gamma$ increases and is even higher than the mIoU without extrapolation at all. The reason, however, is that errors do not accumulate when previous ROIs are not used; when events are sparse (the prerequisite of extrapolation), the ROI naturally moves little, in which case using the previous ROI is the correct decision.

\paragraph{Edge Map} \Fig{fig:miou_edge_sensitivity} compares two ROI prediction networks: one uses the edge map and the other does not, both trained with the same procedure. We show the segmentation accuracy and the speed with and without extrapolation enabled. Variants that use the edge map are consistently more accurate than ones that do not (2\% mIoU difference). Using the edge map has little impact on speed, because the ROI prediction network is very lightweight to begin with.

\Fig{fig:seq_examples} uses a sequence of five consecutive frames to illustrate how the edge map helps correct the ROI drift. The solid and dashed rectangles are the ROIs that are predicted with and without using the edge map, respectively. The first frame is right after the user's eye opens after a blink\no{\footnote{This is why the first edge map is empty; the eye is closed in the previous frame, so the segmentation map and, by extension, the edge map is empty.}}, in which case the event map provides a decent guidance to predict the ROI. However, in the next few frames the eye movements are insignificant, resulting in sparse event maps; in fact, the event map at frame 3 is completely empty. Thus, the ROIs predicted using the event map alone stay almost at the same location. The edge maps during the inactive period, however, still provide decent guidance as they are predicted from the segmentation maps.

\subsection{Evaluation on TEyeD} 
\label{sec:eval:teyed}

Overall, the trend of the results on TEyeD matches that seen in OpenEDS 2020. \Tbl{table: teyed_acc} shows the accuracy and speedup across different methods. Compared to \texttt{RITnet}, \texttt{Ours(S)} achieves a 3.0$\times$ speedup with better accuracy; \texttt{Ours(L)} is the second best in overall accuracy, next to only \texttt{DenseElNet}, which is 4.8$\times$ slower and over 40$\times$ larger in model size (\Tbl{table:model_size}).

By incorporating Auto-ROI and extrapolation, \texttt{Ours(S)+ROI} and \texttt{Ours(S)+ROI+E} achieves a 5.0$\times$ and 5.7$\times$ speedup over \texttt{RITnet}, respectively, while still maintaining a sub-0.7 pixel pupil position error. The results show that our algorithm is robust on the large-scale TEyeD dataset, providing significant speedups over prior methods with higher or competitive accuracy.


\subsection{Failure Cases and Limitations}
\label{sec:eval:fail}

\begin{figure}
\centering
\includegraphics[width=\columnwidth]{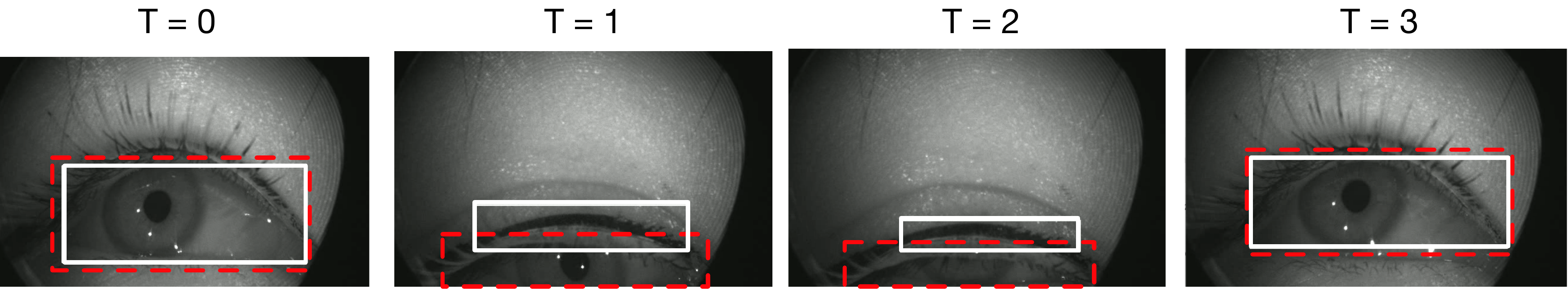}
\caption{A sequence showing failure cases. The solid and dashed boxes are the predicted and ground truth ROIs, respectively.}
\vspace{-10pt}
\label{fig:bad_case}
\end{figure}

We show a failure case using a sequence of four consecutive frames in \Fig{fig:bad_case}. The solid and dashed boxes represent the predicted and ground truth ROIs, respectively. In just four frames, the user's eye has significant vertical eye movements where the eye almost moves out of the camera view and then moves back. The predicted ROIs slightly drifted in the two middle frames. That said, the prediction model is robust enough to correctly predict the ROI after the eye moves back to the view center.

\no{
\subsection{In-Sensor Auto ROI Results}
\label{sec:eval:in_sensor}

\begin{figure}[t]
\centering
\includegraphics[width=\columnwidth]{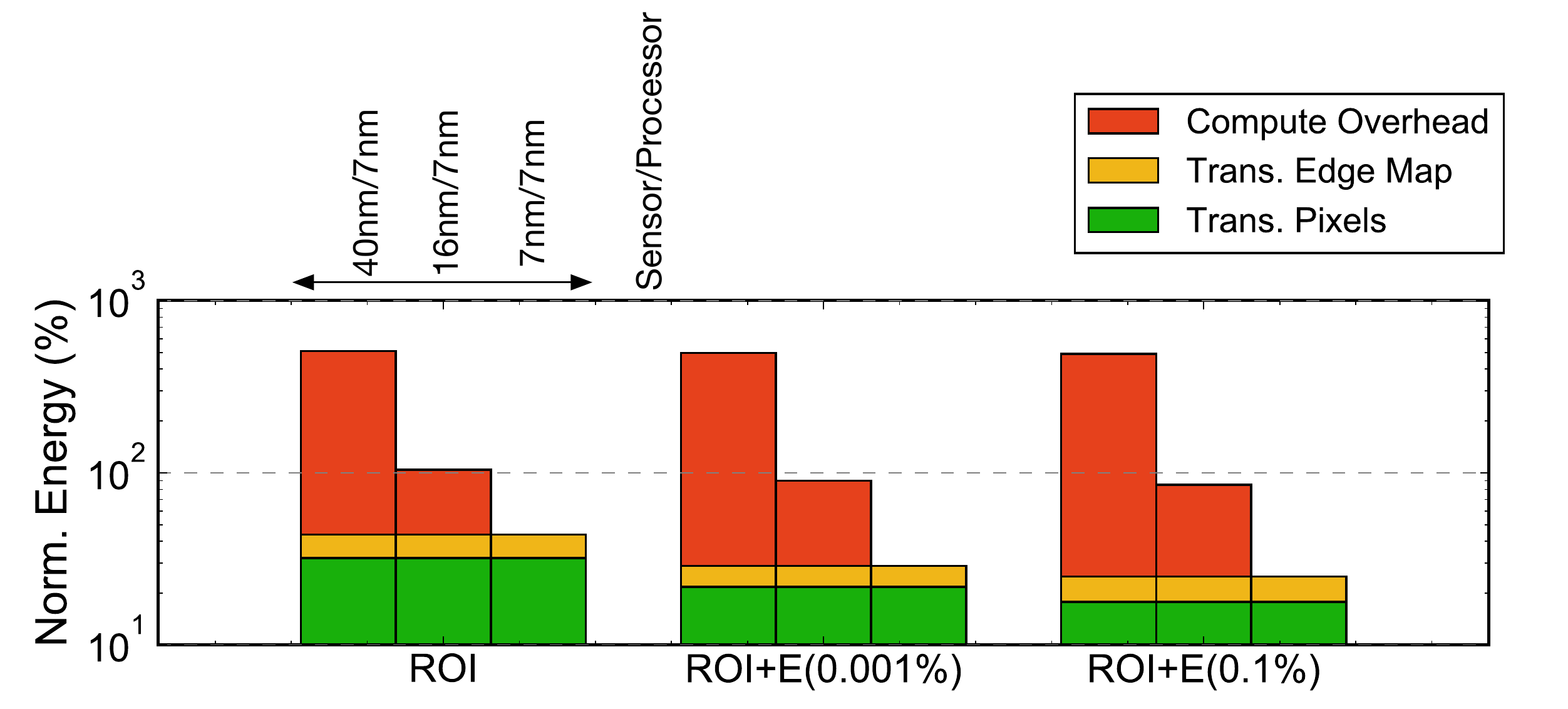}
\caption{The normalized energy across different variants under three sensor/processor technology node combinations. \texttt{Ours(L)} is used as the segmentation network for all the results here.}
\label{fig:savings_overhead}
\end{figure}

\paragraph{Methodology} While compute-capable image sensors such as the Sony IMX 500~\cite{imx500} are on the horizon, most require special partnership and there is no publicly available samples that we are aware of. We use a recent analytical model published by Meta~\cite{liu2019intelligent} to provide a first-order estimation. The energy model estimates the energy consumption of data transmission per byte to be 800 times more than that of computation (i.e., FLOPs/Byte). This energy model matches other recent studies~\cite{kodukula2021rhythmic, han2016eie}. The energy consumption at different process nodes is taken from TSMC's public data~\cite{wu2014advancing, tsmcnode}.

\paragraph{Sensitivity Study} \Fig{fig:savings_overhead} shows how the data transmission energy varies with different sensor-vs-processor process node combinations. We consider three such combinations: 1) an ideal case where both use a 7~nm node, 2) a realistic case where the processor uses a 7~nm node and the sensor uses a 16~nm node, one generation behind, and 3) a less common case where the processor uses a 7~nm node and the sensor uses a 40~nm node, three generations behind.

We show results for three variants, one with the ROI mode only and the other two with ROI and extrapolation. We use two extrapolation variants, \texttt{ROI+E(0.001\%)} and \texttt{ROI+E(0.1\%)}, that correspond to the least and most energy-conserving ones in \Fig{fig:extrapolation_threshold_senstivity}. The numbers in parentheses are the event density threshold. \texttt{Ours(L)} is used as the segmentation network. We normalize the energy with respect to the baseline that always transmits the full-frame images. The figure is a stacked bar plot, where each bar is decoupled into three components: the pixel data transmission energy, the overhead of transmitting the edge map, and the additional computation energy.

When the sensor uses a 40 nm node and the processor uses the advanced 7 nm node (an uncommon combination), the total energy \textit{increases} by about 4 times. This is because the energy overhead of computing inside the sensor on the older process node offsets the energy reduction from reducing the data transmission volume. When the camera sensor uses a more advanced technology node of 16 nm, we start seeing energy reduction. For instance, the total energy is reduced by 14.9\% under \texttt{ROI+E(0.1\%)}.

As the sensor's process node improves over time, the energy saving from in-sensor ROI increases as well. When both the sensor and the processor use a 7 nm node, in-sensor Auto ROI reduces the data transmission energy by 65.6\% even without extrapolation. Extrapolations further reduce the data volume and improves the energy savings. \texttt{ROI+E(0.1\%)} reduces energy by 80.5\%. The result shows that for in-sensor computing to be beneficial, the fabrication technology of the sensor can not lag too much behind that of the processor. Nevertheless, under a realistic 7nm/16nm setting with extrapolation, our system can still reduce the sensor energy by 14.9\%.
}

%% file: sen_fig.tex
\begin{figure}[t]
  \begin{minipage}[t]{0.49\columnwidth}
    \centering
    \includegraphics[width=\columnwidth]{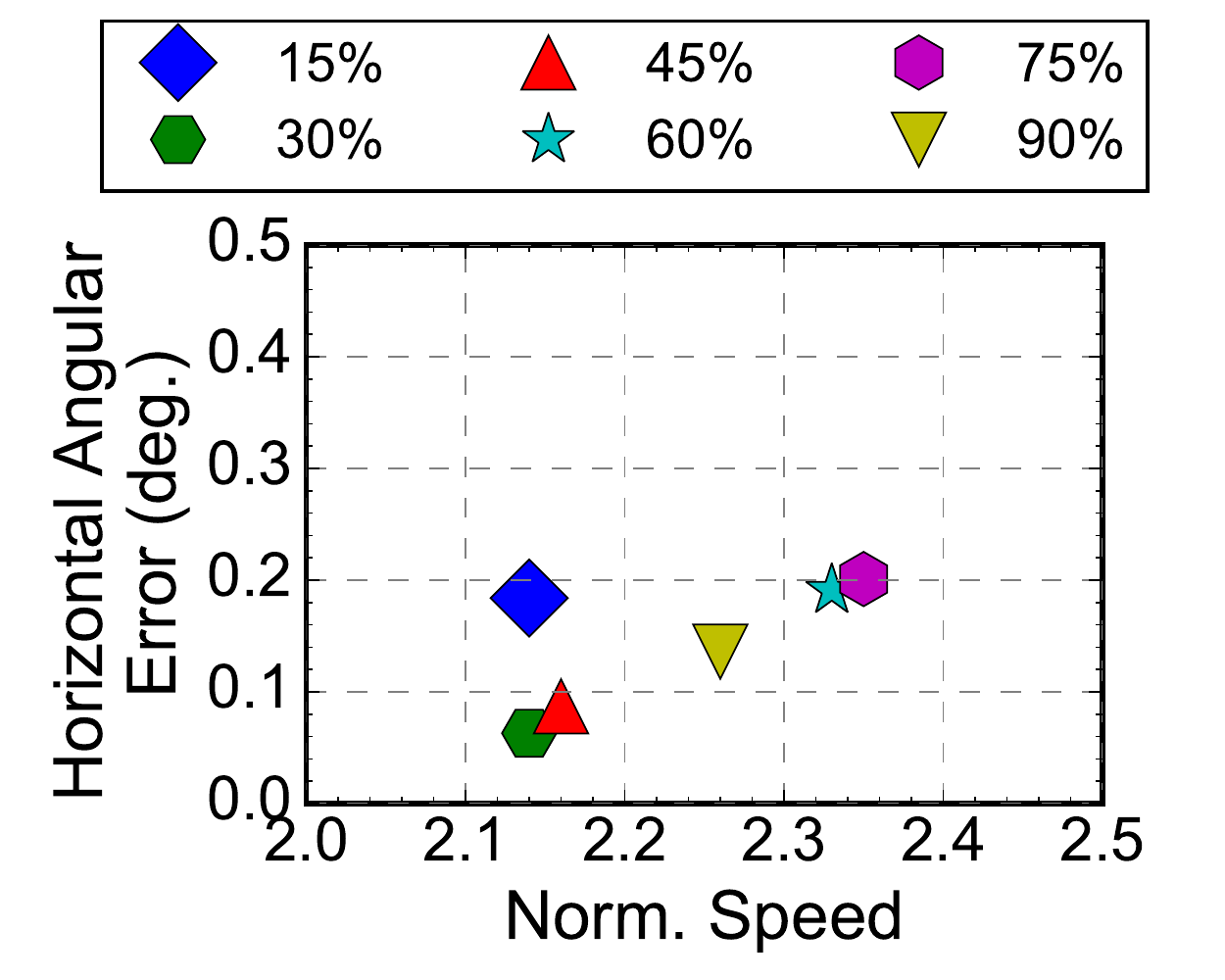}
    \caption{Sensitivity of horizontal gaze error and speedup to different event activation thresholds $\sigma$. All results are from \texttt{Ours(L)+ROI}.}
    \label{fig:event_threshold_sensitivity}
  \end{minipage}
  \hfill
  \begin{minipage}[t]{0.49\columnwidth}
    \centering
    \includegraphics[width=\columnwidth]{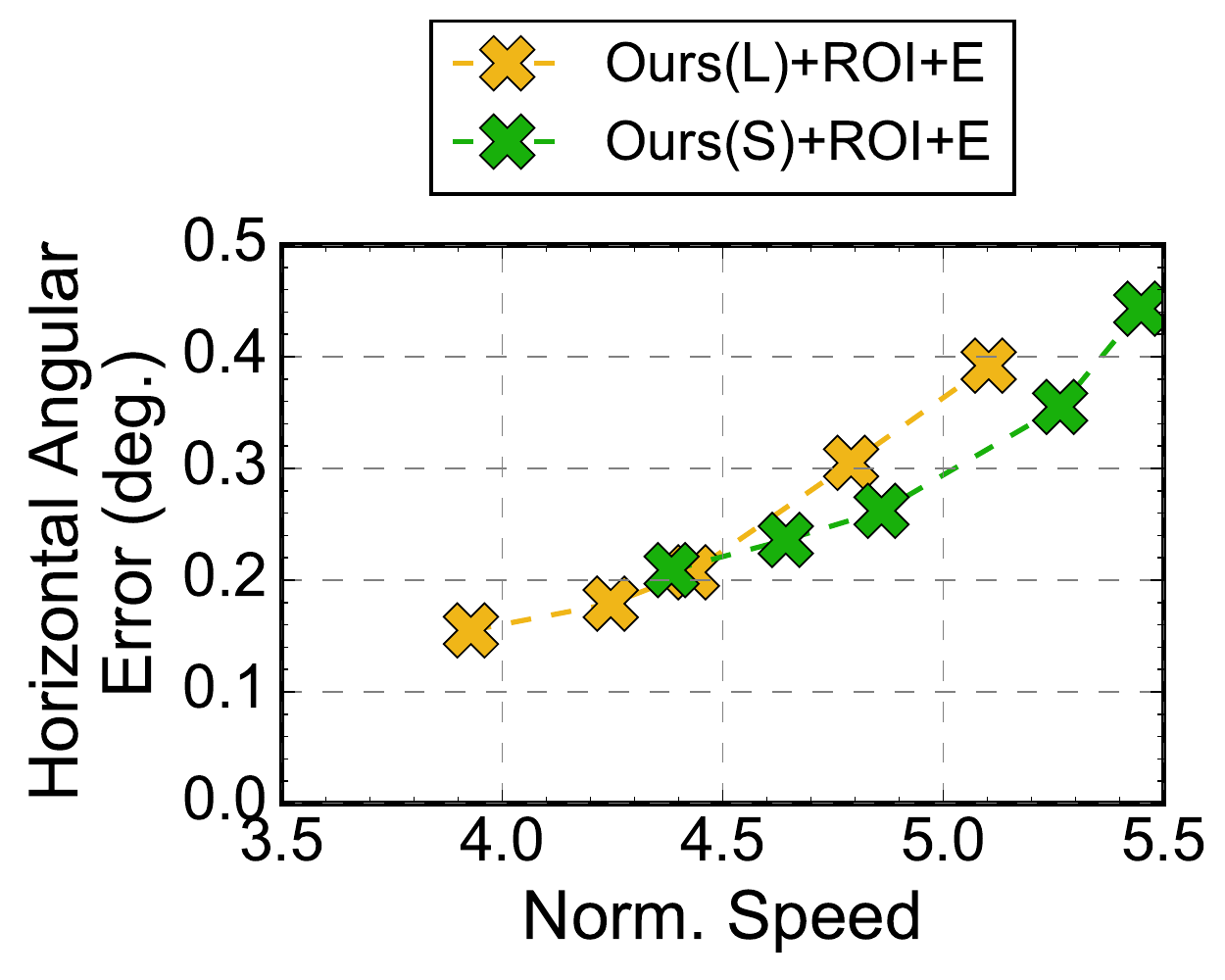}
    \caption{Sensitivity of horizontal gaze error and speedup to different event density thresholds $\gamma$ (from left to right: 0.001\%, 0.005\%, 0.01\%, 0.05\%, 0.1\%).}
    \label{fig:extrapolation_threshold_senstivity}
  \end{minipage}
\end{figure}

%% file: acc_table.tex
\begin{table*}[t]
\caption{Accuracy and speed on TEyeD dataset. The speed is normalized to RITnet. Both the segmentation accuracy (mIoU) and pupil position error in pixel (px) are shown with the standard deviation across the testset.}
\centering
\resizebox{\textwidth}{!}{
\renewcommand*{\arraystretch}{1}
\renewcommand*{\tabcolsep}{2pt}
\begin{tabular}{c c c c c c c c c c c}
\toprule[0.15em]
\textbf{Network} & \textbf{RITnet} & \textbf{DeepVOG} &  \textbf{DenseElNet} & \textbf{Ours (L)} & \textbf{Ours (S)} & \textbf{Ours (L)+ROI}  & \textbf{Ours (S)+ROI} & \textbf{Ours (L)+ROI+E} & \textbf{Ours (S)+ROI+E}  \\
\midrule[0.05em]
mIoU (\%) & 91.70$\pm$3.96 & 94.23$\pm$2.80 & 96.60$\pm$1.76 & 94.94$\pm$2.55 & 93.08$\pm$3.17 & 93.06$\pm$2.48 & 91.81$\pm$3.33 &  92.75$\pm$2.95 & 91.53$\pm$3.77  \\
Pupil Error (px.) & 0.55$\pm$0.17 & 0.49$\pm$0.17 & 0.28$\pm$0.09 & 0.42$\pm$0.13 & 0.50$\pm$0.18 & 0.46$\pm$0.13 & 0.62$\pm$0.25 & 0.49$\pm$0.10 & 0.65$\pm$0.23    \\
\midrule[0.05em]
Norm. Speedup & 1 & 0.74 & 0.40 & 1.93 & 2.97 & 3.87 & 4.95 & 4.86 & 5.69\\
\bottomrule[0.15em]
\end{tabular}
}
\vspace{-10pt}
\label{table: teyed_acc}
\end{table*}

%% file: conc.tex
\section{Conclusion}
\label{sec:conc}

We present an algorithm that enables real-time eye tacking. The algorithm operates at real-time (over 30 Hz on a mobile GPU) --- thanks to our Auto ROI mode, which judiciously processes only the ROI images whenever possible. Our main contribution is a novel ROI prediction algorithm, which emulates an event camera in software and uses events as natural guidance for ROI prediction. While events are largely effective, we show that feedback cues from previous frames are critical to ensure a robust ROI prediction.

\no{
\paragraph{In-Sensor Auto ROI} Our ROI prediction algorithm is lean; with only 30K parameters in size it is small enough to fit in emerging 3D-stacked image sensors, which have MBs of memory (e.g., 8 MB in Sony IMX 500~\cite{imx500}). The ROI prediction algorithm is also extremely lightweight, requiring only 55M FLOPs, while embedded DNN accelerators such as the Arm Ethos-U series execute hundreds of billions of FLOPs per second~\cite{armmicronpu}. These suggest that the ROI prediction algorithm can indeed fit in an image sensor to enable more efficient Auto ROI.


\paragraph{Auto ROI With Event Cameras} It is interesting to explore how an actual event camera can help improve ROI prediction. A real event camera samples temporal events much faster than a conventional camera and thus could potentially improve the ROI prediction.

To leverage the high-frequency samples from event cameras, the downstream processing algorithm must be fast enough to keep up with the event rate. A quick ballpark estimation shows that our ROI prediction algorithm itself can comfortably run at 10 KHz even on an embedded processor, matching the frame rate of event cameras.

\paragraph{General Auto ROI} An interesting research direction is to generalize our ROI prediction algorithm beyond near-eye images and beyond eye tracking. Many tasks in AR/VR such as gesture tracking and SLAM could benefit from ROI-based processing~\cite{kodukula2021rhythmic}. Modern image sensors do provide an ROI mode (``windowing'' in the sensor parlance)~\cite{onseminoip1sn025ka, ov9782}, which, however, relies on users to provide the ROIs. Automatic, robust, and general ROI prediction is still an open question; we are optimistic that the notion of events, along with other feedback cues, will continue to be important in predicting ROIs.
}

%% file: main.bbl
\begin{thebibliography}{10}

\bibitem{armmicronpu}
Arm ethos-u processor series.
\newblock
  \url{https://armkeil.blob.core.windows.net/developer/Files/pdf/arm-ethos-u-processor-series-brief-v2.pdf}.

\bibitem{itohtracker}
Non-dnn eye tracker.
\newblock \url{https://github.com/YutaItoh/3D-Eye-Tracker}.

\bibitem{xavier2dev}
{Nvidia Jetson AGX Xavier}.
\newblock
  \url{https://www.nvidia.com/en-us/autonomous-machines/embedded-systems/jetson-agx-xavier/}.

\bibitem{ov9782}
Omnivision ov9782.
\newblock \url{https://www.ovt.com/sensors/OV9782}.

\bibitem{onseminoip1sn025ka}
Onsemi noip1sn025ka datashee.
\newblock \url{https://www.onsemi.com/pdf/datasheet/noip1sn025ka-d.pdf}.

\bibitem{openeds2019}
Openeds 2019 challenge.
\newblock \url{https://research.fb.com/programs/openeds-challenge/}.

\bibitem{imx500}
{Sony to Release World's First Intelligent Vision Sensors with AI Processing
  Functionality}.
\newblock \url{https://www.sony.com/en/SonyInfo/News/Press/202005/20-037E/}.

\bibitem{tsmcnode}
Tsmc logic technology.
\newblock \url{https://www.tsmc.com/english/dedicatedFoundry/technology/logic}.

\bibitem{akinlar2021accurate}
C.~Akinlar, H.~K. Kucukkartal, and C.~Topal.
\newblock Accurate cnn-based pupil segmentation with an ellipse fit error
  regularization term.
\newblock {\em Expert Systems with Applications}, p. 116004, 2021.

\bibitem{al2013enhanced}
A.~Al-Rahayfeh and M.~Faezipour.
\newblock Enhanced frame rate for real-time eye tracking using circular hough
  transform.
\newblock In {\em 2013 IEEE Long Island Systems, Applications and Technology
  Conference (LISAT)}, pp. 1--6. IEEE, 2013.

\bibitem{amir20183d}
M.~F. Amir and S.~Mukhopadhyay.
\newblock 3d stacked high throughput pixel parallel image sensor with
  integrated reram based neural accelerator.
\newblock In {\em 2018 IEEE SOI-3D-Subthreshold Microelectronics Technology
  Unified Conference (S3S)}, pp. 1--3. IEEE, 2018.

\bibitem{angelopoulos2020event}
A.~N. Angelopoulos, J.~N. Martel, A.~P. Kohli, J.~Conradt, and G.~Wetzstein.
\newblock Event based, near eye gaze tracking beyond 10,000 hz.
\newblock {\em arXiv preprint arXiv:2004.03577}, 2020.

\bibitem{borys2017eye}
M.~Borys and M.~Plechawska-W{\'o}jcik.
\newblock Eye-tracking metrics in perception and visual attention research.
\newblock {\em EJMT}, 3:11--23, 2017.

\bibitem{brown2007automatic}
M.~Brown and D.~G. Lowe.
\newblock Automatic panoramic image stitching using invariant features.
\newblock {\em International journal of computer vision}, 74(1):59--73, 2007.

\bibitem{brunye2019review}
T.~T. Bruny{\'e}, T.~Drew, D.~L. Weaver, and J.~G. Elmore.
\newblock A review of eye tracking for understanding and improving diagnostic
  interpretation.
\newblock {\em Cognitive research: principles and implications}, 4(1):1--16,
  2019.

\bibitem{caligari2013eye}
M.~Caligari, M.~Godi, S.~Guglielmetti, F.~Franchignoni, and A.~Nardone.
\newblock Eye tracking communication devices in amyotrophic lateral sclerosis:
  impact on disability and quality of life.
\newblock {\em Amyotrophic Lateral Sclerosis and Frontotemporal Degeneration},
  14(7-8):546--552, 2013.

\bibitem{canny1986computational}
J.~Canny.
\newblock A computational approach to edge detection.
\newblock {\em IEEE Transactions on pattern analysis and machine intelligence},
  (6):679--698, 1986.

\bibitem{chai2020sensor}
Y.~Chai.
\newblock In-sensor computing for machine vision, 2020.

\bibitem{chandra2015eye}
S.~Chandra, G.~Sharma, S.~Malhotra, D.~Jha, and A.~P. Mittal.
\newblock Eye tracking based human computer interaction: Applications and their
  uses.
\newblock In {\em 2015 International Conference on Man and Machine Interfacing
  (MAMI)}, pp. 1--5. IEEE, 2015.

\bibitem{chaudhary2019ritnet}
A.~K. Chaudhary, R.~Kothari, M.~Acharya, S.~Dangi, N.~Nair, R.~Bailey,
  C.~Kanan, G.~Diaz, and J.~B. Pelz.
\newblock Ritnet: Real-time semantic segmentation of the eye for gaze tracking.
\newblock In {\em 2019 IEEE/CVF International Conference on Computer Vision
  Workshop (ICCVW)}, pp. 3698--3702. IEEE, 2019.

\bibitem{chen20083d}
J.~Chen and Q.~Ji.
\newblock 3d gaze estimation with a single camera without ir illumination.
\newblock In {\em 2008 19th International Conference on Pattern Recognition},
  pp. 1--4. IEEE, 2008.

\bibitem{chollet2017xception}
F.~Chollet.
\newblock Xception: Deep learning with depthwise separable convolutions.
\newblock In {\em Proceedings of the IEEE conference on computer vision and
  pattern recognition}, pp. 1251--1258, 2017.

\bibitem{clay2019eye}
V.~Clay, P.~K{\"o}nig, and S.~Koenig.
\newblock Eye tracking in virtual reality.
\newblock {\em Journal of Eye Movement Research}, 12(1), 2019.

\bibitem{damianeye}
G.~Damian, T.~Delbruck, and P.~Lichtsteiner.
\newblock Eye tracking using event-based silicon retina.

\bibitem{dennard1974design}
R.~H. Dennard, F.~H. Gaensslen, H.-N. Yu, V.~L. Rideout, E.~Bassous, and A.~R.
  LeBlanc.
\newblock Design of ion-implanted mosfet's with very small physical dimensions.
\newblock {\em IEEE Journal of Solid-State Circuits}, 9(5):256--268, 1974.

\bibitem{dierkes2019fast}
K.~Dierkes, M.~Kassner, and A.~Bulling.
\newblock A fast approach to refraction-aware eye-model fitting and gaze
  prediction.
\newblock In {\em Proceedings of the 11th ACM Symposium on Eye Tracking
  Research \& Applications}, pp. 1--9, 2019.

\bibitem{eki20219}
R.~Eki, S.~Yamada, H.~Ozawa, H.~Kai, K.~Okuike, H.~Gowtham, H.~Nakanishi,
  E.~Almog, Y.~Livne, G.~Yuval, et~al.
\newblock 9.6 a 1/2.3 inch 12.3 mpixel with on-chip 4.97 tops/w cnn processor
  back-illuminated stacked cmos image sensor.
\newblock In {\em 2021 IEEE International Solid-State Circuits Conference
  (ISSCC)}, vol.~64, pp. 154--156. IEEE, 2021.

\bibitem{el1999pixel}
A.~El~Gamal, D.~X. Yang, and B.~A. Fowler.
\newblock Pixel-level processing: why, what, and how?
\newblock In {\em Sensors, Cameras, and Applications for Digital Photography},
  vol. 3650, pp. 2--13. International Society for Optics and Photonics, 1999.

\bibitem{ester1996density}
M.~Ester, H.-P. Kriegel, J.~Sander, X.~Xu, et~al.
\newblock A density-based algorithm for discovering clusters in large spatial
  databases with noise.
\newblock In {\em kdd}, vol.~96, pp. 226--231, 1996.

\bibitem{feng2019asv}
Y.~Feng, P.~Whatmough, and Y.~Zhu.
\newblock Asv: Accelerated stereo vision system.
\newblock In {\em Proceedings of the 52nd Annual IEEE/ACM International
  Symposium on Microarchitecture}, pp. 643--656, 2019.

\bibitem{fischler1981random}
M.~A. Fischler and R.~C. Bolles.
\newblock Random sample consensus: a paradigm for model fitting with
  applications to image analysis and automated cartography.
\newblock {\em Communications of the ACM}, 24(6):381--395, 1981.

\bibitem{fuhl2021teyed}
W.~Fuhl, G.~Kasneci, and E.~Kasneci.
\newblock Teyed: Over 20 million real-world eye images with pupil, eyelid, and
  iris 2d and 3d segmentations, 2d and 3d landmarks, 3d eyeball, gaze vector,
  and eye movement types.
\newblock {\em arXiv preprint arXiv:2102.02115}, 2021.

\bibitem{fuhl2016pupilnet}
W.~Fuhl, T.~Santini, G.~Kasneci, and E.~Kasneci.
\newblock Pupilnet: Convolutional neural networks for robust pupil detection.
\newblock {\em arXiv preprint arXiv:1601.04902}, 2016.

\bibitem{fuhl2016else}
W.~Fuhl, T.~C. Santini, T.~K{\"u}bler, and E.~Kasneci.
\newblock Else: Ellipse selection for robust pupil detection in real-world
  environments.
\newblock In {\em Proceedings of the Ninth Biennial ACM Symposium on Eye
  Tracking Research \& Applications}, pp. 123--130, 2016.

\bibitem{gallego2019event}
G.~Gallego, T.~Delbruck, G.~Orchard, C.~Bartolozzi, B.~Taba, A.~Censi,
  S.~Leutenegger, A.~Davison, J.~Conradt, K.~Daniilidis, et~al.
\newblock Event-based vision: A survey.
\newblock {\em arXiv preprint arXiv:1904.08405}, 2019.

\bibitem{garbin2019openeds}
S.~J. Garbin, Y.~Shen, I.~Schuetz, R.~Cavin, G.~Hughes, and S.~S. Talathi.
\newblock Openeds: Open eye dataset.
\newblock {\em arXiv preprint arXiv:1905.03702}, 2019.

\bibitem{girshick2015fast}
R.~Girshick.
\newblock Fast r-cnn.
\newblock In {\em Proceedings of the IEEE international conference on computer
  vision}, pp. 1440--1448, 2015.

\bibitem{girshick2014rich}
R.~Girshick, J.~Donahue, T.~Darrell, and J.~Malik.
\newblock Rich feature hierarchies for accurate object detection and semantic
  segmentation.
\newblock In {\em Proceedings of the IEEE conference on computer vision and
  pattern recognition}, pp. 580--587, 2014.

\bibitem{han2016eie}
S.~Han, X.~Liu, H.~Mao, J.~Pu, A.~Pedram, M.~A. Horowitz, and W.~J. Dally.
\newblock Eie: Efficient inference engine on compressed deep neural network.
\newblock {\em ACM SIGARCH Computer Architecture News}, 44(3):243--254, 2016.

\bibitem{hansen2009eye}
D.~W. Hansen and Q.~Ji.
\newblock In the eye of the beholder: A survey of models for eyes and gaze.
\newblock {\em IEEE transactions on pattern analysis and machine intelligence},
  32(3):478--500, 2009.

\bibitem{hansen2005eye}
D.~W. Hansen and A.~E. Pece.
\newblock Eye tracking in the wild.
\newblock {\em Computer Vision and Image Understanding}, 98(1):155--181, 2005.

\bibitem{haruta20174}
T.~Haruta, T.~Nakajima, J.~Hashizume, T.~Umebayashi, H.~Takahashi,
  K.~Taniguchi, M.~Kuroda, H.~Sumihiro, K.~Enoki, T.~Yamasaki, et~al.
\newblock 4.6 a 1/2.3 inch 20mpixel 3-layer stacked cmos image sensor with
  dram.
\newblock In {\em 2017 IEEE International Solid-State Circuits Conference
  (ISSCC)}, pp. 76--77. IEEE, 2017.

\bibitem{he2017mask}
K.~He, G.~Gkioxari, P.~Doll{\'a}r, and R.~Girshick.
\newblock Mask r-cnn.
\newblock In {\em Proceedings of the IEEE international conference on computer
  vision}, pp. 2961--2969, 2017.

\bibitem{hennessey2006single}
C.~Hennessey, B.~Noureddin, and P.~Lawrence.
\newblock A single camera eye-gaze tracking system with free head motion.
\newblock In {\em Proceedings of the 2006 symposium on Eye tracking research \&
  applications}, pp. 87--94, 2006.

\bibitem{howard2017mobilenets}
A.~G. Howard, M.~Zhu, B.~Chen, D.~Kalenichenko, W.~Wang, T.~Weyand,
  M.~Andreetto, and H.~Adam.
\newblock Mobilenets: Efficient convolutional neural networks for mobile vision
  applications.
\newblock {\em arXiv preprint arXiv:1704.04861}, 2017.

\bibitem{hu2020gaze}
Z.~Hu.
\newblock Gaze analysis and prediction in virtual reality.
\newblock In {\em 2020 IEEE Conference on Virtual Reality and 3D User
  Interfaces Abstracts and Workshops (VRW)}, pp. 543--544. IEEE, 2020.

\bibitem{huang2017densenet}
G.~Huang, Z.~Liu, L.~Van Der~Maaten, and K.~Q. Weinberger.
\newblock Densely connected convolutional networks.
\newblock In {\em Proceedings of the IEEE conference on computer vision and
  pattern recognition}, pp. 4700--4708, 2017.

\bibitem{itoh2014interaction}
Y.~Itoh and G.~Klinker.
\newblock Interaction-free calibration for optical see-through head-mounted
  displays based on 3d eye localization.
\newblock In {\em 2014 IEEE symposium on 3d user interfaces (3dui)}, pp.
  75--82. IEEE, 2014.

\bibitem{jacob2003eye}
R.~J. Jacob and K.~S. Karn.
\newblock Eye tracking in human-computer interaction and usability research:
  Ready to deliver the promises.
\newblock In {\em The mind's eye}, pp. 573--605. Elsevier, 2003.

\bibitem{jianfeng2014eye}
L.~Jianfeng and L.~Shigang.
\newblock Eye-model-based gaze estimation by rgb-d camera.
\newblock In {\em Proceedings of the IEEE Conference on Computer Vision and
  Pattern Recognition Workshops}, pp. 592--596, 2014.

\bibitem{kar2017review}
A.~Kar and P.~Corcoran.
\newblock A review and analysis of eye-gaze estimation systems, algorithms and
  performance evaluation methods in consumer platforms.
\newblock {\em IEEE Access}, 5:16495--16519, 2017.

\bibitem{kim2016real}
H.~Kim, S.~Leutenegger, and A.~J. Davison.
\newblock Real-time 3d reconstruction and 6-dof tracking with an event camera.
\newblock In {\em European Conference on Computer Vision}, pp. 349--364.
  Springer, 2016.

\bibitem{kim2019eye}
S.-H. Kim, G.-S. Lee, H.-J. Yang, et~al.
\newblock Eye semantic segmentation with a lightweight model.
\newblock In {\em 2019 IEEE/CVF International Conference on Computer Vision
  Workshop (ICCVW)}, pp. 3694--3697. IEEE, 2019.

\bibitem{kodukula2021rhythmic}
V.~Kodukula, A.~Shearer, V.~Nguyen, S.~Lingutla, Y.~Liu, and R.~LiKamWa.
\newblock Rhythmic pixel regions: multi-resolution visual sensing system
  towards high-precision visual computing at low power.
\newblock In {\em Proceedings of the 26th ACM International Conference on
  Architectural Support for Programming Languages and Operating Systems}, pp.
  573--586, 2021.

\bibitem{kong2016hypernet}
T.~Kong, A.~Yao, Y.~Chen, and F.~Sun.
\newblock Hypernet: Towards accurate region proposal generation and joint
  object detection.
\newblock In {\em Proceedings of the IEEE conference on computer vision and
  pattern recognition}, pp. 845--853, 2016.

\bibitem{kothari2021ellseg}
R.~S. Kothari, A.~K. Chaudhary, R.~J. Bailey, J.~B. Pelz, and G.~J. Diaz.
\newblock Ellseg: An ellipse segmentation framework for robust gaze tracking.
\newblock {\em IEEE Transactions on Visualization and Computer Graphics},
  27(5):2757--2767, 2021.

\bibitem{kumagai20181}
O.~Kumagai, A.~Niwa, K.~Hanzawa, H.~Kato, S.~Futami, T.~Ohyama, T.~Imoto,
  M.~Nakamizo, H.~Murakami, T.~Nishino, et~al.
\newblock A 1/4-inch 3.9 mpixel low-power event-driven back-illuminated stacked
  cmos image sensor.
\newblock In {\em 2018 IEEE International Solid-State Circuits
  Conference-(ISSCC)}, pp. 86--88. IEEE, 2018.

\bibitem{kwon2020low}
M.~Kwon, S.~Lim, H.~Lee, I.-S. Ha, M.-Y. Kim, I.-J. Seo, S.~Lee, Y.~Choi,
  K.~Kim, H.~Lee, et~al.
\newblock A low-power 65/14nm stacked cmos image sensor.
\newblock In {\em 2020 IEEE International Symposium on Circuits and Systems
  (ISCAS)}, pp. 1--4. IEEE, 2020.

\bibitem{li2018etracker}
B.~Li, H.~Fu, D.~Wen, and W.~Lo.
\newblock Etracker: A mobile gaze-tracking system with near-eye display based
  on a combined gaze-tracking algorithm.
\newblock {\em Sensors}, 18(5):1626, 2018.

\bibitem{liu20204}
C.~Liu, L.~Bainbridge, A.~Berkovich, S.~Chen, W.~Gao, T.-H. Tsai, K.~Mori,
  R.~Ikeno, M.~Uno, T.~Isozaki, et~al.
\newblock A 4.6 $\mu$m, 512$\times$ 512, ultra-low power stacked digital pixel
  sensor with triple quantization and 127db dynamic range.
\newblock In {\em 2020 IEEE International Electron Devices Meeting (IEDM)}, pp.
  16--1. IEEE, 2020.

\bibitem{liu2019intelligent}
C.~Liu, A.~Berkovich, S.~Chen, H.~Reyserhove, S.~S. Sarwar, and T.-H. Tsai.
\newblock Intelligent vision systems--bringing human-machine interface to
  ar/vr.
\newblock In {\em 2019 IEEE International Electron Devices Meeting (IEDM)}, pp.
  10--5. IEEE, 2019.

\bibitem{loh2007processor}
G.~H. Loh, Y.~Xie, and B.~Black.
\newblock Processor design in 3d die-stacking technologies.
\newblock {\em Ieee Micro}, 27(3):31--48, 2007.

\bibitem{lowe1999object}
D.~G. Lowe.
\newblock Object recognition from local scale-invariant features.
\newblock In {\em Proceedings of the seventh IEEE international conference on
  computer vision}, vol.~2, pp. 1150--1157. Ieee, 1999.

\bibitem{lu2014adaptive}
F.~Lu, Y.~Sugano, T.~Okabe, and Y.~Sato.
\newblock Adaptive linear regression for appearance-based gaze estimation.
\newblock {\em IEEE transactions on pattern analysis and machine intelligence},
  36(10):2033--2046, 2014.

\bibitem{mania2021gaze}
K.~Mania, A.~McNamara, and A.~Polychronakis.
\newblock Gaze-aware displays and interaction.
\newblock In {\em ACM SIGGRAPH 2021 Courses}, pp. 1--67. 2021.

\bibitem{mathur2021dynamic}
P.~Mathur, T.~Mittal, and D.~Manocha.
\newblock Dynamic graph modeling of simultaneous eeg and eye-tracking data for
  reading task identification.
\newblock In {\em ICASSP 2021-2021 IEEE International Conference on Acoustics,
  Speech and Signal Processing (ICASSP)}, pp. 1250--1254. IEEE, 2021.

\bibitem{mele2012gaze}
M.~L. Mele and S.~Federici.
\newblock Gaze and eye-tracking solutions for psychological research.
\newblock {\em Cognitive processing}, 13(1):261--265, 2012.

\bibitem{mudassar2019camera}
B.~A. Mudassar, P.~Saha, Y.~Long, M.~F. Amir, E.~Gebhardt, T.~Na, J.~H. Ko,
  M.~Wolf, and S.~Mukhopadhyay.
\newblock A camera with brain--embedding machine learning in 3d sensors.
\newblock In {\em 2019 Design, Automation \& Test in Europe Conference \&
  Exhibition (DATE)}, pp. 680--685. IEEE, 2019.

\bibitem{mukhopodhyay2018camel}
S.~Mukhopodhyay, M.~Wolf, M.~F. Amir, E.~Gebahrdt, J.~H. Ko, J.~H. Kung, and
  B.~A. Musassar.
\newblock The camel approach to stacked sensor smart cameras.
\newblock In {\em 2018 Design, Automation \& Test in Europe Conference \&
  Exhibition (DATE)}, pp. 1299--1303. IEEE, 2018.

\bibitem{palmero2020openeds2020}
C.~Palmero, A.~Sharma, K.~Behrendt, K.~Krishnakumar, O.~V. Komogortsev, and
  S.~S. Talathi.
\newblock Openeds2020: open eyes dataset.
\newblock {\em arXiv preprint arXiv:2005.03876}, 2020.

\bibitem{patney2016perceptually}
A.~Patney, J.~Kim, M.~Salvi, A.~Kaplanyan, C.~Wyman, N.~Benty, A.~Lefohn, and
  D.~Luebke.
\newblock Perceptually-based foveated virtual reality.
\newblock In {\em ACM SIGGRAPH 2016 Emerging Technologies}, pp. 1--2. 2016.

\bibitem{plopski2016automated}
A.~Plopski, J.~Orlosky, Y.~Itoh, C.~Nitschke, K.~Kiyokawa, and G.~Klinker.
\newblock Automated spatial calibration of hmd systems with unconstrained
  eye-cameras.
\newblock In {\em 2016 IEEE International Symposium on Mixed and Augmented
  Reality (ISMAR)}, pp. 94--99. IEEE, 2016.

\bibitem{ramesh2018long}
B.~Ramesh, S.~Zhang, Z.~W. Lee, Z.~Gao, G.~Orchard, and C.~Xiang.
\newblock Long-term object tracking with a moving event camera.
\newblock In {\em Bmvc}, p. 241, 2018.

\bibitem{ren2015faster}
S.~Ren, K.~He, R.~Girshick, and J.~Sun.
\newblock Faster r-cnn: Towards real-time object detection with region proposal
  networks.
\newblock {\em Advances in neural information processing systems}, 28:91--99,
  2015.

\bibitem{ronneberger2015unet}
O.~Ronneberger, P.~Fischer, and T.~Brox.
\newblock U-net: Convolutional networks for biomedical image segmentation.
\newblock In {\em International Conference on Medical image computing and
  computer-assisted intervention}, pp. 234--241. Springer, 2015.

\bibitem{strandvall2009eye}
T.~Strandvall.
\newblock Eye tracking in human-computer interaction and usability research.
\newblock In {\em IFIP Conference on Human-Computer Interaction}, pp. 936--937.
  Springer, 2009.

\bibitem{swirski2012robust}
L.~{\'S}wirski, A.~Bulling, and N.~Dodgson.
\newblock Robust real-time pupil tracking in highly off-axis images.
\newblock In {\em Proceedings of the symposium on eye tracking research and
  applications}, pp. 173--176, 2012.

\bibitem{Swirski2013}
L.~\'Swirski and N.~A. Dodgson.
\newblock A fully-automatic, temporal approach to single camera, glint-free 3d
  eye model fitting [abstract].
\newblock In {\em Proceedings of ECEM 2013}, Aug. 2013.

\bibitem{tsugawa2017pixel}
H.~Tsugawa, H.~Takahashi, R.~Nakamura, T.~Umebayashi, T.~Ogita, H.~Okano,
  K.~Iwase, H.~Kawashima, T.~Yamasaki, D.~Yoneyama, et~al.
\newblock Pixel/dram/logic 3-layer stacked cmos image sensor technology.
\newblock In {\em 2017 IEEE International Electron Devices Meeting (IEDM)}, pp.
  3--2. IEEE, 2017.

\bibitem{venezia20181}
V.~C. Venezia, A.~C.-W. Hsiung, K.~Ai, X.~Zhao, Z.~Lin, D.~Mao, A.~Yazdani,
  E.~A. Webster, and L.~A. Grant.
\newblock 1.5um dual conversion gain, backside illuminated image sensor using
  stacked pixel level connections with 13ke-full-well capacitance and 0.8
  e-noise.
\newblock In {\em 2018 IEEE International Electron Devices Meeting (IEDM)}, pp.
  10--1. IEEE, 2018.

\bibitem{wang2021edge}
Z.~Wang, Y.~Zhao, Y.~Liu, and F.~Lu.
\newblock Edge-guided near-eye image analysis for head mounted displays.
\newblock In {\em 2021 IEEE International Symposium on Mixed and Augmented
  Reality (ISMAR)}, pp. 11--20. IEEE, 2021.

\bibitem{weikersdorfer2013simultaneous}
D.~Weikersdorfer, R.~Hoffmann, and J.~Conradt.
\newblock Simultaneous localization and mapping for event-based vision systems.
\newblock In {\em International Conference on Computer Vision Systems}, pp.
  133--142. Springer, 2013.

\bibitem{whitmire2016eyecontact}
E.~Whitmire, L.~Trutoiu, R.~Cavin, D.~Perek, B.~Scally, J.~Phillips, and
  S.~Patel.
\newblock Eyecontact: scleral coil eye tracking for virtual reality.
\newblock In {\em Proceedings of the 2016 ACM International Symposium on
  Wearable Computers}, pp. 184--191, 2016.

\bibitem{wood2016learning}
E.~Wood, T.~Baltru{\v{s}}aitis, L.-P. Morency, P.~Robinson, and A.~Bulling.
\newblock Learning an appearance-based gaze estimator from one million
  synthesised images.
\newblock In {\em Proceedings of the Ninth Biennial ACM Symposium on Eye
  Tracking Research \& Applications}, pp. 131--138, 2016.

\bibitem{wu2014advancing}
S.-Y. Wu, C.~Lin, S.~Yang, J.~Liaw, and J.~Cheng.
\newblock Advancing foundry technology with scaling and innovations.
\newblock In {\em Proceedings of Technical Program-2014 International Symposium
  on VLSI Technology, Systems and Application (VLSI-TSA)}, pp. 1--3. IEEE,
  2014.

\bibitem{yiu2019deepvog}
Y.-H. Yiu, M.~Aboulatta, T.~Raiser, L.~Ophey, V.~L. Flanagin, P.~Zu~Eulenburg,
  and S.-A. Ahmadi.
\newblock Deepvog: Open-source pupil segmentation and gaze estimation in
  neuroscience using deep learning.
\newblock {\em Journal of neuroscience methods}, 324:108307, 2019.

\bibitem{yu201914nm}
D.~Yu, C.~jae Lee, M.~Park, J.~Park, S.~Hwang, J.~Lee, S.~Yu, H.~Shin, B.~Kim,
  J.-W. Choi, et~al.
\newblock 14nm finfet process technology platform for over 100m pixel density
  and ultra low power 3d stack cmos image sensor.
\newblock In {\em 2019 IEEE International Electron Devices Meeting (IEDM)}, pp.
  8--1. IEEE, 2019.

\bibitem{zhang2017eye}
X.~Zhang, X.~Liu, S.-M. Yuan, and S.-F. Lin.
\newblock Eye tracking based control system for natural human-computer
  interaction.
\newblock {\em Computational intelligence and neuroscience}, 2017, 2017.

\bibitem{zhang2019evaluation}
X.~Zhang, Y.~Sugano, and A.~Bulling.
\newblock Evaluation of appearance-based methods and implications for
  gaze-based applications.
\newblock In {\em Proceedings of the 2019 CHI Conference on Human Factors in
  Computing Systems}, pp. 1--13, 2019.

\bibitem{zhang2015appearance}
X.~Zhang, Y.~Sugano, M.~Fritz, and A.~Bulling.
\newblock Appearance-based gaze estimation in the wild.
\newblock In {\em Proceedings of the IEEE conference on computer vision and
  pattern recognition}, pp. 4511--4520, 2015.

\bibitem{zhang2017mpiigaze}
X.~Zhang, Y.~Sugano, M.~Fritz, and A.~Bulling.
\newblock Mpiigaze: Real-world dataset and deep appearance-based gaze
  estimation.
\newblock {\em IEEE transactions on pattern analysis and machine intelligence},
  41(1):162--175, 2017.

\bibitem{zhu2018euphrates}
Y.~Zhu, A.~Samajdar, M.~Mattina, and P.~Whatmough.
\newblock Euphrates: Algorithm-soc co-design for low-power mobile continuous
  vision.
\newblock {\em arXiv preprint arXiv:1803.11232}, 2018.

\bibitem{zhu2005eye}
Z.~Zhu and Q.~Ji.
\newblock Eye gaze tracking under natural head movements.
\newblock In {\em 2005 IEEE Computer Society Conference on Computer Vision and
  Pattern Recognition (CVPR'05)}, vol.~1, pp. 918--923. IEEE, 2005.

\end{thebibliography}
